\documentclass[twocolumn,a4paper,showpacs,superscriptaddress,pra,eqsecnum]{revtex4}

\usepackage{amsfonts,amsmath,graphicx,color,epsfig}
\DeclareGraphicsExtensions{.jpeg,.jpg,.ps,.eps,.pdf}
\newcommand{\beq}{\begin{equation}}
\newcommand{\eeq}{\end{equation}}
\newcommand{\bqa}{\begin{eqnarray}}
\newcommand{\eqa}{\end{eqnarray}}
\newcommand{\nn}{\nonumber}
\newcommand{\nl}[1]{\nn \\ && {#1}\,}

\newcommand{\erf}[1]{Eq.~(\ref{#1})}

\newcommand{\ea}{{\it et al.~}}

\newcommand{\dg}{^\dagger}

\newcommand{\smallfrac}[2]{\mbox{$\frac{#1}{#2}$}}
\newcommand{\half}{\smallfrac{1}{2}}
\newcommand{\bra}[1]{\langle{#1}|}
\newcommand{\ket}[1]{|{#1}\rangle}

\newcommand{\sch}{Schr\"odinger}

\newcommand{\sq}[1]{\left[ {#1} \right]}
\newcommand{\cu}[1]{\left\{ {#1} \right\}}
\newcommand{\ro}[1]{\left( {#1} \right)}

\newcommand{\tr}[1]{{\rm Tr}\sq{ {#1} }}

\definecolor{nblue}{rgb}{0.2,0.2,0.7}
\definecolor{ngreen}{rgb}{0.2,0.6,0.2}
\definecolor{nred}{rgb}{0.7,0.2,0.2}
\definecolor{nyellow}{rgb}{0.7,0.6,0.2}
\definecolor{npurple}{rgb}{0.7,0.1,0.7}
\definecolor{nbackground}{rgb}{1,1,1}
\definecolor{ngrey}{rgb}{0.5,0.5,0.5}
\definecolor{nbrown}{rgb}{0.6,0.4,0.2}
\definecolor{nblack}{rgb}{0,0,0}

\newcommand{\blk}{\color{nblack}}

\begin{document}

\title{Entanglement and Symmetry: A Case Study in \\ Superselection Rules, Reference Frames, and Beyond}
  \author{S. J. Jones}
\affiliation{Centre for Quantum Computer Technology, Centre for
  Quantum Dynamics, School of Science, Griffith University, Brisbane,
  4111 Australia}
\author{H. M. Wiseman}
\email{H.Wiseman@griffith.edu.au}
\affiliation{Centre for Quantum Computer Technology, Centre for
  Quantum Dynamics, School of Science, Griffith University, Brisbane,
  4111 Australia}
\author{S. D. Bartlett}
\affiliation{School of Physics, The University of Sydney, Sydney,
New South Wales 2006, Australia}
  \author{J. A. Vaccaro}
\affiliation{Centre for Quantum Computer Technology, Centre for
  Quantum Dynamics, School of Science, Griffith University, Brisbane,
  4111 Australia}
  \author{D. T. Pope}
\affiliation{Centre for Quantum Computer Technology, Centre for
  Quantum Dynamics, School of Science, Griffith University, Brisbane,
  4111 Australia}
  \date{October 20, 2006}

\begin{abstract}
In recent years it has become apparent that constraints on
possible quantum operations, such as those constraints imposed by
superselection rules (SSRs), have a profound effect on quantum
information theoretic concepts like bipartite entanglement. This
paper concentrates on a particular example: the constraint that
applies when the parties (Alice and Bob) cannot distinguish among
certain quantum objects they have. This arises naturally in the
context of ensemble quantum information processing such as in
liquid NMR. We discuss how a SSR for the symmetric group can be
applied, and show how the extractable entanglement can be
calculated analytically in certain cases, with a maximum bipartite
entanglement in an ensemble of $N$ Bell-state pairs scaling as
$\log(N)$ as $N \to \infty$. We discuss the apparent disparity
with the asymptotic ($N \to \infty)$ recovery of unconstrained
entanglement for other sorts of superselection rules, and show
that the disparity disappears when the correct notion of applying
the symmetric group SSR to multiple copies is used. Next we
discuss reference frames in the context of this SSR, showing the
relation to the work of von Korff and Kempe [Phys. Rev. Lett. {\bf
93}, 260502 (2004)]. The action of a reference frame can be
regarded as the analog of activation in mixed-state entanglement.
We also discuss the analog of distillation:  there exist states
such that one copy can act as an imperfect reference frame for
another copy.  Finally we present an example of a stronger
operational constraint, that operations must be non-collective as
well as symmetric. Even under this stronger constraint we
nevertheless show that Bell-nonlocality (and hence entanglement)
can be demonstrated for an ensemble of $N$ Bell-state pairs no
matter how large $N$ is. This last work is a generalization of
that of Mermin [Phys. Rev. D {\bf 22}, 356 (1980)].
   \end{abstract}


\maketitle

\section{Introduction}\label{Intro}

The entanglement of disjoint (typically spatially separate)
quantum systems is at the heart of quantum information processing
~\cite{NieChu00}.  For bipartite pure states under LOCC (local
operations and classical communication) the quantification and
transformation of entanglement is now well understood. However, it
is also now well understood that the non-ideal situation of mixed
states, which pertains in practice, is far more complicated (or
richer, to put a different spin on it) \cite{Hor01}. In recent
years it has also become apparent that a situation, 
in which only certain operations can be performed, also leads to an
interesting theory of entanglement, even if the states are pure. One
approach, leading to a generalized notion of entanglement, dismisses
altogether with the bipartite setting \cite{Barnum03,Barnum04}. A
less radical, and more obviously applicable, idea is to restrict the
local operations to those that are invariant under a superselection
rule (SSR)
\cite{Ver03,WisVac03,BarWis03,KMP04,SchVerCir04,BarDohSpeWis06,VacAnsWisJac06}.
At the same time, the nature of quantum reference frames in the
bipartite setting has also been hotly debated (see for example
Refs.~\cite{RudSan01,EnkFuc02a,SanBarRudKni03,Wis04}).

Much of the work in this area
\cite{Ver03,WisVac03,SchVerCir04,BarDohSpeWis06,RudSan01,
EnkFuc02a,SanBarRudKni03,Wis04} has concentrated upon the case of
a U(1)-SSR. This is the SSR that can be motivated by considering
the conservation of a locally additive scalar quantity with a
discrete spectrum \cite{Wis04}. It can also be applied to quantum
optics experiments which lack an optical phase reference (that is,
which lack a shared clock of sufficient precision)
\cite{SanBarRudKni03}. Many simplifications arise from this SSR
because U(1) is Abelian (there is only one generator,
corresponding to the local operator of the conserved quantity).
Non-Abelian Lie-group SSRs (with non-commuting generators) have
also been considered \cite{KMP04} but relatively little attention
has been paid to SSRs arising from discrete groups. An example
with obvious application to ensemble quantum information
processing is the symmetric group $S_N$ (the group of permutations
of $N$ objects) \cite{BarWis03}.

This paper explores issues in entanglement under operations
constrained by symmetry. We use the $S_N$-SSR formalism of
Ref.~\cite{BarWis03}, but also go beyond that work. This work is
important for a number of reasons. First, as noted above, the
symmetric group has been relatively neglected in studies of
entanglement constrained by a SSR. For the U(1)-SSR concepts like
bound entanglement (of two distinct types), activation, and
distillation have been shown to apply, in analogy to these
concepts in mixed-state entanglement. Although not immediately
obvious, we construct specific examples to show how these concepts
apply to the $S_N$-SSR. Second, we clarify the notion of a
reference frame for the $S_N$ group, linking in with the work of
von Korff and Kempe \cite{vKK04}. Finally, we give an example
where it is not obvious that the symmetry constraints on the
system can be formulated as a SSR. 
Nevertheless we show that, even
under such constraints, it is possible to exhibit Bell-nonlocality
\cite{Bel64} for an ensemble of identically prepared singlets.

\section{Entanglement and SSRs}

\subsection{Concepts of Entanglement} \label{sec:concepts}

The term entanglement was coined by \sch\ \cite{Sch35} as the
property that bipartite pure states have when they are not product
states. \sch\ showed that for such an entangled state, one party
(say Bob) could, via a measurement on his system, collapse Alice's
system with some probability to {\em any} state vector (except those
in the null space of Alice's reduced state matrix). \sch\ thought
this nonlocality was unreasonable enough to be called a ``paradox''
\cite{Sch36}. A generation later, Bell \cite{Bel64} discovered that
such states had an even stronger form of nonlocality: for certain
measurement schemes, the correlations between the results of Alice
and Bob cannot be explained by any locally causal theory. This
property, which we will call Bell-nonlocality, we regard as the
strongest operational notion of entanglement.

\subsubsection{Separability and Local Preparability}
When correlations in mixed states were first studied in earnest
\cite{Wer89}, it became clear that the question as to whether a
state was entangled was no longer straightforward. In particular,
Werner showed that there were nonseparable states such that the
measurement correlations of Alice and Bob could nevertheless be
explained by a local theory involving hidden variables. Nonseparable
states are states that cannot be written in the form \bqa
\rho &=& 
\sum_k \wp_k |\psi_k\rangle\langle \psi_k| \otimes |\phi_k\rangle\langle\phi_k|\nonumber\\
&\equiv& \biguplus_k  \sqrt{\wp_k} \ro{ \ket{\psi_k}\otimes \ket{\phi_k} } \nonumber\\
&\equiv& \uplus \sqrt{\wp_1} \ro{ \ket{\psi_1}\otimes \ket{\phi_1} }
\uplus \sqrt{\wp_2} \ro{ \ket{\psi_2}\otimes \ket{\phi_2} } \uplus
\cdots  \nl{} \eqa Here, following Ref.~\cite{VacAnsWisJac06}, we
have defined a notation that we will use throughout this paper, that
for an arbitrary {\em ray} $\ket{r}$, we have $\uplus \ket{r} \equiv
+ \ket{r}\bra{r}$. Werner called nonseparable states ``EPR
correlated states''. They are sometimes identified with
``entangled'' states but we will call them non-locally-preparable states. This name captures the physical significance of
such states: they cannot be prepared by LOCC from a product state.

\subsubsection{$n$-Distillability and Bound
Entanglement}\label{Distillable} Since Werner, the 
richness of the entanglement of mixed states has been further
developed, involving concepts such as bound entanglement,
distillation, $n$-distillability, and activation \cite{Hor01}.
Here, following Ref.~\cite{BarDohSpeWis06}, we concentrate upon
those properties of mixed state entanglement for which there are
obvious analogs in pure state entanglement constrained by SSRs.
First, as noted above, it is useful to define the class of \emph{locally preparable}
states, which are those states that are preparable from a product state
using LOCC. Another useful class is the class of states that are
\emph{distillable}~\cite{Ben96}. States in the distillable class are such
that $n$ copies can be converted into $nr$ pure maximally
entangled states via LOCC for some $r>0$ in the limit
$n\to\infty$. A \emph{pure} state is either locally preparable or distillable,
depending on whether it is a product state or not. On the other
hand, there are mixed states that are neither locally preparable nor distillable. These
are the so-called \emph{bound entangled} states~\cite{Hor98}.

For mixed states, deciding whether a state is locally preparable is known to be
an NP-hard problem computationally \cite{Gur02}, but algorithms to
do so exist \cite{DohParSpe02}.  It is not known if it is even
possible to determine whether a state is distillable. For this reason, a
related, but simpler to characterize, class has been defined: the
states that are \emph{1-distillable}~\cite{Div00,Dur00}. A
state $\rho$ is 1-distillable if by LOCC Alice and Bob can, with some
probability, create from it a non-separable two-qubit state. (Note
that for two qubits, there are no bound entangled
states~\cite{Hor97}.) By extension, a state $\rho$ is
\emph{$n$-distillable} if $\rho^{\otimes n}$ is 1-distillable. (If a
state is $n$-distillable for some $n$ then it is distillable.) Thus, the set of distillable states includes the 1-distillable states,
and in fact it has recently been shown that the $n$-distillable states are a subset of the distillable states
$\forall\ n$~\cite{Wat04}.  Since the 1-distillable states are a subset of distillable states,
there are clearly mixed states that are neither locally preparable nor
1-distillable.  We shall refer to these states as being \emph{1-bound}.

Note that although a nonseparable two-qubit state is always
distillable, this does not mean that undistilled copies can be used
to demonstrate Bell-nonlocality, as Werner showed \cite{Wer89}.
However, in our  work pertaining to SSRs, when we demonstrate that a
state is 1-distillable, we do this by showing that it is possible for Alice
and Bob by LOCC to create with some probability a {\em pure}
entangled state, as this is strictly stronger than the requirements
for being 1-distillable. Thus, for these purposes, a state that is 1-distillable
allows Alice and Bob to demonstrate Bell-nonlocality, which, as
noted above, we regard as the strongest notion of entanglement.

\subsubsection{Closing the gap: PPT-Channels}\label{Become}

Returning to the 1-bound states in general, this class can be divided
into two by considering what would happen if we were to give Alice
and Bob a PPT-channel. That is, a channel that can distribute only
bipartite states for which the partial transpose is positive. 
This allows Alice and Bob
to perform PPT-operations as well as LOCC. A PPT operation is one
that preserves the positivity of the partial (with respect to Alice
or Bob) transpose of states~\cite{Rai01}. With this addition, Alice
and Bob can locally prepare all states with a positive partial
transpose, which includes some states which are 1-bound \cite{Hor98}. We will
call these the bound states that \emph{become locally preparable}.
Conversely, the rest of the 1-bound states, those that are not PPT,
become 1-distillable under LOCC plus all PPT operations
\cite{Egg01}. Hence we call this class (which is also non-empty
\cite{Wat04}) \emph{become 1-distillable}.

\subsubsection{Activation}

Physically, a PPT-channel is equivalent to 
supplying Alice and Bob with an infinite number of copies of every
state in the become locally preparable class.
Access to these states automatically makes them locally
preparable. However, it is not necessary to use all of the states to
make 1-distillable a state in become 1-distillable. Rather, for every $\rho$ in
become 1-distillable there exists a state $\sigma$ in 
become locally preparable
such that
$\rho \otimes \sigma$ is 1-distillable. This is known as {\em
activation}~\cite{Hor99}. Note the distinction from {\em
distillation}, in which for some $\rho$ which is in become 1-distillable, 
there exists an $n$ such that $\rho^{\otimes n}$ is 1-distillable
\cite{Wat04}. Note that it is trivially the case that any state
$\rho$ which can become locally preparable does so 
given a suitable state $\sigma$ which can become locally preparable: 
one simply chooses $\sigma = \rho$.

\subsubsection{Measures of Entanglement}

Finally for this section, we define some {\em measures} of
entanglement. The entanglement of formation, $E_F$, of a mixed state
$\rho$ is the minimum ratio, in the asymptotic limit, of the number
of singlets used to the number of copies of $\rho$ created thereby,
using LOCC \cite{HillWoot97}. Similarly, the distillable
entanglement, $E_D$, is the asymptotic yield of arbitrarily pure
singlets that can be prepared by LOCC from copies of $\rho$
\cite{BennetDiVSmolWoot96}. By definition, both of these measures
are {\em partially additive}. That is, $n$ copies of a state $\rho$
contains $n$ times the entanglement of a single copy;
$E(\rho^{\otimes n})=nE(\rho)$. Also by definition \cite{Hor99}, and
the fact that LOCC cannot increase entanglement, $E_F$ is an upper
bound on $E_D$. In general it is a strict upper bound, which is
obvious from the existence of bound entangled states where $E_F \neq
0$ when $E_D = 0$. However for pure states $E_F = E_D$. Since we
will be concerned with states that can be made pure by LOCC, there
is no need to distinguish between $E_F$ and $E_D$. For a bipartite
pure state $\ket{\Psi}$ the entanglement (measured in e-bits
\cite{BennetDiVSmolWoot96}) is defined as the von Neumann entropy of
either subsystem's reduced density matrix, \beq
E(\ket{\Psi})=-\tr{\rho_A\log_2\rho_A}=-\tr{\rho_B\log_2\rho_B},
\eeq where the reduced density matrices for Alice and Bob are
defined as $\rho_A={\rm Tr_{B}}[\ket{\Psi}\bra{\Psi}]$ and
$\rho_B={\rm Tr_{A}}[\ket{\Psi}\bra{\Psi}]$ respectively. ${\rm
Tr_{A,B}}$ signifies the partial trace operation with respect to
Alice or Bob.

\subsection{Superselection Rules}\label{sec:SSRs}

\subsubsection{SSRs as an Operational Restriction}

Originally \cite{Wic52}, SSRs were regarded as restrictions on the
states that a system can be in. This could be restated
operationally, 
as a restriction on the means of preparing a system. Since any
operation could be part of a system-preparation procedure, it is
only sensible to say that a SSR is a restriction on the operations
that can be performed on a system. For an SSR for charge (the
first
such SSR ever proposed) 
\cite{Wic52}, this restriction would amount to saying that it is not
possible to create superpositions of different charge eigenstates.
Alternatively, all operations on the system must commute with
charge-preserving operations such as measurement of charge.
Charge-preserving operations can be built up from transformations in
the Lie group U(1) generated by the charge operator. This
formulation allows the concept of SSRs to be generalized to
arbitrary compact Lie groups, or finite groups
\cite{BarWis03,KMP04}, as we now explain.

The SSR for a group $G$ of physical transformations can be defined
operationally as follows. Consider for the moment a single party,
Alice, who possesses a quantum system, described by a Hilbert
space $\mathbb{H}_A$. Let the physical transformation
corresponding to an element $g$ of $G$ be denoted $\hat T_A(g)$.
Then the $G$-SSR  is the rule that all operations must be
$G$-invariant. That is, if ${\cal O}$ is the completely positive
map $\rho \to {\cal O}{\rho}$ representing the operation, then
\beq \forall \rho \textrm{ and } \forall g \in G\,, \;{\cal
O}[\hat T_A(g) \rho \hat T\dg_A(g)] = \hat T_A(g)[{\cal
O}\rho]\hat T\dg_A(g). \eeq Note that ``operations" includes
unitaries, where ${\cal O}\rho = \hat{U}\rho\hat{U}\dg$, and also
measurements, where for example ${\cal O}_r\rho =
\hat{M}_r\rho\hat{M}_r\dg$ and $\sum_r \hat{M}_r\dg \hat{M}_r =
\hat{1}$.

According to this definition, we would say that a SSR for charge
$\hat{Q}_A$, for example, would be a SSR for the group U(1)
generated by $\hat{Q}_A$. Such a SSR can be motivated from
consideration of a conservation law for global charge $\hat{Q}_A$.
Note however that we do not assume that the operational
restriction described by a general SSR must be derivable from a
conservation law. For the purposes of this paper, it is more
fruitful to regard a $G$-SSR as being due to the lack (by Alice)
of an appropriate {\em reference frame}
\cite{AhaSus67,BarRudSpe05,BarDohSpeWis06}.
This idea will be
explored later in the particular context of the $S_N$-SSR.

\subsubsection{SSRs and Mixing} \label{SSRmixing}

All quantum information processing ultimately ends in measurement.
If a $G$-SSR is in force over the entire process, then no outcomes
will be changed if the state matrix for the quantum system $\rho$ is
replaced by the state matrix $\hat T_A(g)\rho \hat T\dg_A(g)$ for
any $g \in G$. That is, under the $G$-SSR the state of the quantum
system is represented by an equivalence class of state matrices. The
{\em most mixed} state matrix to which $\rho$ is physically
equivalent is \beq \mathcal{G}_A\rho \equiv  |G|^{-1} \sum_{g \in G}
\hat T_A(g) \rho \hat T\dg_A(g) \eeq for finite groups, where $|G|$
is the group order, and \beq \mathcal{G}_A\rho \equiv \int_{{}_{G}}
\text{d}\mu(g)\, \hat T_A(g) \rho \hat T\dg_A(g) \eeq for compact
Lie  groups, where $\text{d}\mu(g)$ is the Haar measure. We call
this the $G$-invariant state, as \beq \label{AlocalGinv} \forall g
\in G, \; \hat T_A(g) [\mathcal{G}_A\rho] \hat T\dg_A(g) =
\mathcal{G}_A\rho. \eeq For traditional SSRs, i.e. groups with a
single generator $\hat{Q}_A=\sum_q q\, \hat\Pi_q $, the
$G$-invariant state is simply the block-diagonal state ${\cal G}_A
\rho = \sum_q \, \hat\Pi_q \, \rho \, \hat\Pi_q.$

This maximum-entropy member of the equivalence class is the one
containing no irrelevant information, and hence it is the natural
representation of the state of the system as a state matrix. This
state can also be given an operational interpretation
\cite{VacAnsWisJac06}. Given the $D$-dimensional quantum system
with state $\rho$ and a heat bath at temperature $T$, work can be
extracted by allowing the system to come to thermal equilibrium.
The maximum amount of extractable work is $k_{\rm B}T[\log D -
S(\rho)]$, where $S$ is the von Neumann entropy \cite{OppHor02}.
Under the constraint of a $G$-SSR, the amount of extractable work
is reduced by (the positive quantity) $k_{\rm B}T \Delta_G(\rho)$,
where $\Delta_G(\rho) = S({\cal G}_A\rho)-S(\rho)$ is precisely
the amount of ``irrelevant information'' in $\rho$.

It is very important to note that if Alice has two systems with
states $\rho_1$ and $\rho_2$, such that ${\cal G}_A(\rho_1\otimes
\rho_2)$ equals \beq  |G|^{-1}
  \sum_{g \in G}
  [\hat T_1(g)\otimes \hat T_2(g) ] \rho [\hat T_1\dg(g)\otimes \hat T_2\dg(g)
  ],
\eeq then this state is not the same as  ${\cal G}_A\rho_1 \otimes
{\cal G}_A \rho_2$, which equals \beq |G|^{-2}  \sum_{g,g' \in G}
  [\hat T_1(g)\otimes \hat T_2(g') ] \rho [\hat T_1\dg(g)\otimes \hat T_2\dg(g') ] .
\eeq That is why in the above we have referred to {\em the} quantum
system, not {\em a} quantum system. If we are considering the whole
quantum system (or at least all parts to which the SSR applies),
then the state $\rho$ of the system can be replaced by ${\cal
G}_A\rho$. But if there are other quantum systems that may enter
into the quantum information processing at a later time, then it is
{\em not} true in general that ${\cal G}_A\rho$ contains all of the
relevant information about that system.

\subsubsection{Bipartite SSRs}
In this paper we are concerned with the impact of SSRs on
entanglement, rather than extractable work (although the latter
is, in the bipartite setting, also related to entanglement
\cite{VacAnsWisJac06}). In this context we have to define the
concept of {\em local} SSRs. That is, the local operations of
Alice and Bob (say) must respect local SSRs, rather than a global
SSR. This is obviously applicable in the case when a SSR is
motivated by a conservation law for a locally additive quantity.
It is also applicable more generally if Alice and Bob each lack a
reference frame. It turns out that for the purpose of non-local
quantum information processing, what is important is that Alice
and Bob have a shared reference frame. Furthermore, such a
reference frame need
only be correlated between the two parties. This point will be
clarified by later examples.

For the concept of a local SSR or local reference frame to make
sense, the physical transformation on the joint Hilbert space
$\mathbb{H}_A\otimes\mathbb{H}_B$ corresponding to an element $g$ of
the group $G$ must have the following form: \beq \hat T(g) = \hat
T_A(g)\otimes \hat T_B(g). \eeq Now if Alice and Bob lack reference
frames, then the effective state for the bipartite system is the
locally $G$-invariant state \cite{BarWis03} \beq
 ({\cal G}_A \otimes {\cal G}_B) \rho,
 \eeq
where ${\cal G}_A$ is defined as above, and ${\cal G}_B$
similarly, and these act locally according to the tensor-product
structure of the joint system. Note that this state is in general
very different from the globally $G$-invariant state
 \beq
 {\cal G}\rho = \sum |G|^{-1}
  \sum_{g \in G}
  [\hat T_A(g) \otimes \hat T_B(g)] \rho [\hat T_A\dg (g) \otimes \hat T_B\dg
  (g)].
 \eeq
Just as in the case of a single party, it is important to remember
that $\rho$ can be replaced by $({\cal G}_A \otimes {\cal G}_B)
\rho$ only if it is the state of the entire quantum system shared
by Alice and Bob (or at least all parts to which the SSR applies).

\subsubsection{SSRs and Hilbert Space
(Technicalities)}\label{Technicalities}

To determine the effect of SSRs on entanglement it is necessary to
understand how a SSR induces a structure on Hilbert space. A local
$G$-SSR for Alice splits $\mathbb{H}^A$ into ``charge sectors''
labeled by $y$:
\begin{equation}
    \mathbb{H}^A = \bigoplus_y \mathbb{H}^A_y ,
\end{equation}
where each $\mathbb{H}^A_y$ carries inequivalent representations
$\hat T^A_y$ of $G$. The sectors are then further decomposed into
tensor products:
\begin{equation}
  \mathbb{H}^A_y = \mathbb{M}^A_{y} \otimes \mathbb{Q}^A_y .
\end{equation}
This is technically known as dividing the system into subsystems.
The subsystem $\mathbb{M}^A_y$ carries an irreducible representation
(irrep) $\hat t^A_y(g)$ and the subsystem $\mathbb{Q}^A_y$ carries a
trivial representation of $G$. That is to say, \beq \hat T^A_y(g) =
\hat t^A_y(g) \otimes \hat{I}^A_y. \eeq For an Abelian SSR such as
charge, the subsystems $\mathbb{M}^A_y$ are one-dimensional, and so
the additional tensor product structure within the irreps is not
required. However, for a non-Abelian SSR such as we will consider
later, they are nontrivial.

The subsystems $\mathbb{Q}^A_y$ are clearly $G$-invariant. They have
been called noiseless subsystems, or decoherence-free subsystems,
relative to the decoherence map $\mathcal{G}_A$~\cite{Kni00}. By
contrast, the subsystems $\mathbb{M}^A_y$ become completely mixed
under the action of $\mathcal{G}_A$, because $\hat t^A_y(g)$ is
irreducible. Thus the action of $\mathcal{G}_A$ on an arbitrary
state matrix $\rho$ is, in terms of this decomposition,
\begin{equation}
  \mathcal{G}_A\rho = \sum_y \mathcal{D}^A_{y} \otimes
  \mathcal{I}^A_{y}(\hat\Pi^A_y \rho \hat\Pi^A_y ).
\end{equation}
Here $\hat{\Pi}^A_y$ is the projection onto the charge sector $y$,
$\mathcal{D}^A_{y}$ is the trace-preserving map that takes every
operator for the subsystem $\mathbb{M}^A_y$ to a maximally mixed
operator (i.e. proportional to the identity operator on that space),
and $\mathcal{I}^A_{y}$ is the identity map over operators for the
subsystem $\mathbb{Q}^A_y$.  The effect of the local superselection
rule, then, is to remove the coherence between different local
charge sectors (as in the Abelian case) {\em and} to make the
subsystems $\mathbb{M}^A_y$ completely mixed.  The same structure
arises for $\mathbb{H}^B$ and provides an analogous decomposition of
$\mathcal{G}_B$. For further details, see~\cite{BarWis03,KMP04}.

\subsection{Concepts of Entanglement Constrained by
SSRs}\label{sec:EntSSR} In this section we summarize the results of
Ref.~\cite{BarDohSpeWis06}, showing the analogies between
mixed-state entanglement and pure-state entanglement constrained by
a SSR. The various concepts of entanglement explored in
Sec.~\ref{sec:concepts} arise from considering two parties able to
perform LOCC. Adding the constraint of a local $G$-SSR (that is,
that the local operations must be $G$-invariant) we say that the two
parties can perform $G$-LOCC.

\subsubsection{Local Preparability}

The class of pure bipartite states that are locally preparable
under $G$-LOCC will call \emph{${G\text{-SSR}}$ locally preparable}. 
Just as preparable under LOCC means preparable from states that are local
(separable), so preparable under $G$-LOCC means preparable from
states that respect the $G$-SSR (i.e. that are locally
$G$-invariant). It is trivial to see that a pure bipartite state
$\ket{\psi}$ is ${G\text{-SSR}}$ locally preparable 
iff (i) the state is a product state, and (ii) it is locally
$G$-invariant. Note that not all pure product states are
${G\text{-SSR}}$ locally preparable; it is a {\em
smaller} class than the locally preparable states.

\subsubsection{$n$-Distillability and Bound Entanglement}

The class of pure states that are 1-distillable under $G$-LOCC,
which we call \emph{${G\text{-SSR}}$ 1-distillable},
 is defined as those states
$\ket{\psi}$ for which the following is true: The two parties can,
by local measurements, project $\ket{\psi}$ onto a
$2{\times}2$-dimensional subspace with nonzero probability, such
that the projected state is (i) locally $G$-invariant and (ii)
non-separable. The significance of the first condition is that the
SSR is now irrelevant, so that the usual condition (nonseparability)
is all that is required for 1-distillability. It is not difficult to
see  \cite{BarWis03} that $|\psi\rangle$ is ${G\text{-SSR}}$ 1-distillable
iff $\mathcal{G}_A \otimes \mathcal{G}_B[|\psi\rangle\langle\psi |]$
is 1-distillable under LOCC.

Both the class of ${G\text{-SSR}}$ locally preparable and ${G\text{-SSR}}$ 1-distillable states are non-empty in general
(i.e. for a general SSR). Moreover,
as with mixed-state entanglement, there is a proper gap between
these two classes.  The class of states in the gap contains both
product and non-product pure states, and is analogous to the class
of 1-bound states in mixed-state entanglement.

The concepts of $n$-distillability with the SSR constraint (and the
corresponding classes of pure states, \emph{${G\text{-SSR}}\ n$-distillable})  
can be defined analogously to the case of unconstrained entanglement. It is
not difficult to illustrate the phenomenon of distillation; that is,
to find examples of states that are ${G\text{-SSR}}$ distillable but not
 ${G\text{-SSR}}$ 1-distillable \cite{KMP04}. Here ${G\text{-SSR}}\ {\rm distillable}=
{G\text{-SSR}}\ \infty$-distillable is the class of distillable pure states
under this constraint.

\subsubsection{Closing the gap}

Just as in mixed-state entanglement adding a PPT channel removes
the 1-bound class, so it is possible to augment $G$-LOCC in such a way
that any pure state in the gap between ${G\text{-SSR}}$ locally preparable and
${G\text{-SSR}}$ 1-distillable becomes either locally preparable or
1-distillable.  In this case the augmentation is very simple: one
simply lifts the restriction of the local SSR by providing Alice
and Bob with a shared reference frame.
With this additional resource, Alice
and Bob can now implement any operation in LOCC.

Augmenting $G$-LOCC to LOCC divides the proper gap of pure states
between ${G\text{-SSR}}$ locally preparable and ${G\text{-SSR}}$
1-distillable into two classes, both of which are non-empty.  All
product states that are not locally $G$-invariant (i.e., product
states not in ${G\text{-SSR}}$ locally preparable) \emph{become}
locally preparable with $G$-LOCC plus the shared reference frame
for $G$.  We call this class \emph{${G\text{-SSR}}$ become locally
preparable}.  This result follows directly from the fact that all
pure product states are locally preparable with unrestricted LOCC.
Similarly all non-product pure states which are not in
${G\text{-SSR}}$ 1-distillable \emph{become} 1-distillable under
$G$-LOCC plus the shared reference frame for $G$.  We thus call
this class \emph{${G\text{-SSR}}$ become 1-distillable}.  This
result follows directly from the fact that all pure non-product
states are 1-distillable with unrestricted LOCC.

\subsubsection{Activation}

Again, just as in the mixed-state case, it is not necessary to
completely lift the SSR constraint in order to make any particular
state $\ket{\psi}$ either ${G\text{-SSR}}$ locally preparable or
${G\text{-SSR}}$ 1-distillable. Rather, all that is needed is some other pure
state $\ket{\phi}$ which is ${G\text{-SSR}}$ become locally preparable. Again, this is trivial
if $\ket{\psi}$ is ${G\text{-SSR}}$ become locally preparable; one simply chooses
$\ket{\phi}=\ket{\psi}$. But the result is nontrivial when
$\ket{\psi}$ is ${G\text{-SSR}}$ locally preparable, and says that a state
$\ket{\phi}$ which is ${G\text{-SSR}}$ become locally preparable exists such that
$\ket{\phi}\otimes\ket{\psi} \in\  {G\text{-SSR}}$ 1-distillable. This is
analogous to activation and is an example of a partial reference
frame.

\subsubsection{Measures of Entanglement}
As discussed in Sec.~\ref{SSRmixing}, in the unipartite setting a
SSR in general reduces the maximum work that can be extracted from
a system, and that is quantified by the $G$-invariant state.
Similarly, in the bipartite setting the amount of entanglement
that can be extracted from a system under $G$-LOCC is less than
under LOCC, and the locally $G$-invariant state again quantifies
this reduction. The extractable entanglement \footnote{In Ref.
\cite{BarWis03} they actually term this the \emph{entanglement
constrained by a superselection rule}.}
from a single copy is given by \cite{BarWis03}\beq
E_{G\textrm{-SSR}}(\rho) = E_D[({\cal G}_A \otimes {\cal G}_B)
\rho]. \eeq As noted earlier, there is no way known to compute the
distillable entanglement for a general mixed state. Thus we will
restrict our attention to cases \cite{VacAnsWisJac06} where it is
identical to the entanglement of formation. Also note that if a
state (mixed or otherwise) can be used to demonstrate
Bell-nonlocality then it necessarily has nonzero extractable
entanglement.

\section{The Symmetric Group SSR}

\subsection{The constraint of symmetry}

The importance of symmetry as a constraint becomes apparent when
dealing with many identical systems, that is, ensembles. By the
term \emph{ensemble quantum information processing} we mean: (i)
there are $N$ (typically $\gg 1$) identical ``molecules'' each
consisting of $M$ ``atoms'' (typically qubits); (ii) all
operations are {\em symmetric} (i.e. affect each molecule
identically).

For example, in a nuclear magnetic resonance (NMR) experiment
\cite{NMRQIP} each molecule contains $M$ atoms typically having a
spin-$\frac12$ nucleus. Operations may be implemented using radio
frequency (RF) magnetic pulses and an antenna. For the case of
$M=4$ the qubits could be the spin-$\frac12$ nuclei of ${}^1$H,
${}^{17}$O, ${}^{13}$C, ${}^{19}$F. Another example occurs in spin
squeezing experiments \cite{Ueda}. In this case each molecule is a
single two-level, or multi-level, atom ($M=1$). Operations are
implemented using uniform laser fields (and detectors for them),
and thus affect all molecules identically.

In NMR quantum information processing it is also the case that the
molecules are typically prepared in highly mixed states, and the
detection efficiency is very small. These are practical constraints
that apply to current experimental techniques rather than
fundamental constraints such as those previously studied as SSRs.
The consequences of such practical constraints will be discussed
later (the first of these can be overcome at least for small
molecules \cite{Anw04}).

There are $N!$ possible permutations of $N$ molecules. The set of
these permutations $p$ (under the permutation operation) form the
{\em symmetric group} $S_N$. The fact that symmetric operations
must affect the identical molecules in the same way leads to what
is known as the $S_N$-SSR. Another way of stating this is to say
that only symmetric operations can be performed on ensemble
quantum information processing systems.

Using the SSR formalism of Ref. \cite{BarWis03}, the restriction
on operations ${\cal O}$ for ensemble quantum information
processing systems can be stated as \beq \mathcal{O}[\hat T(p)
\rho \hat T^\dag(p)] = \hat T(p)[ \mathcal{O} \rho]\hat
T^\dag(p),\forall p\in S_N,\eeq where $p$ is a permutation of the
$N$ molecules and $\hat T(p)$ is the unitary operator that
implements that permutation. The $N$ molecules can each be thought
of as subsystems of $M$ atoms (e.g. for $M=4$, the $N$ subsystems
could be made up of a $\ {}^{1}$H atom,  $\ {}^{17}$O atom, $\
{}^{13}$C atoms, and $\ {}^{19}$F atom). Each of the atoms within
a subsystems is acted on by the same $\hat T(p)$, because they are
attached to the same molecule.

When the $S_N$-SSR is in effect the allowable operations on the
system are restricted to being symmetric.  Under such operations
the state $\rho$ is indistinguishable from the states $\hat T(p)
\rho \hat T^\dag(p)$ for any $p\in S_N$. Thus we define the most
mixed state with which $\rho$ is equivalent (the $S_N$-invariant
or randomly permuted state) as \beq \mathcal{P} \rho \equiv
\frac{1}{N!} \sum_{p \in S_N} \hat T(p) \rho \hat T^\dag(p). \eeq
Under the $S_N$-SSR it is operationally appropriate to use
$\mathcal{P}\rho$ to describe the state $\rho$.

\subsection{Local $S_N$-SSR}


NMR quantum information processing with pure states may allow the
possibility of scalable quantum computing. In this paper we are
not concerned with this question, but rather a question of
principle: even with pure states, is there entanglement between
different subsystems comprising atoms of the same species? Say we
can create molecules such that there is entanglement between two
species of atom (call them $A$ and $B$) on each molecule, as in
Ref. \cite{Anw04}. Then if we could isolate an individual
molecule, and give one of the relevant atoms to Alice and the
other to Bob, then Alice and Bob would share \blk entanglement. We
could even ``give'' one atom ($A$) to Alice and one ($B$) to Bob
without splitting the molecule, merely by saying that Alice can
control an applied magnetic field and antenna resonant with the
frequency of $A$'s nucleus, and Bob similarly with $B$'s nucleus.
However, the symmetry constraint means that Alice and Bob cannot
isolate a single molecule. So the question then becomes: what is
the nature of the entanglement between Alice's ensemble of $A$
atoms and Bob's ensemble of $B$ atoms?

Both Alice and Bob are restricted from individually
addressing the $N$ molecules in their possession, 
so we must apply the $S_N$-SSR locally. That is to say, the
effective quantum state is $\left({\cal P}_A \otimes {\cal
P}_B\right) \rho$. To understand this, it is helpful to consider a
simple example; say $M=3$ (nuclei $A$, $A'$ and $B$, per molecule)
and $N=2$ (there are two molecules, 1 and 2). We consider that the
$A$s and $A'$s belong to Alice and the $B$s to Bob. The typical
situation in NMR is to assume that the two molecules are prepared
identically. However, for illustrative purposes it will be useful
to consider the following state, where the molecules are not
prepared identically: \beq \ket{\psi} =  \ket{\uparrow_A^1
\uparrow_{A'}^1 \uparrow_B^1}\ket{\downarrow_A^2
\downarrow_{A'}^2\downarrow_B^2}.\eeq This is so that we can allow
for (and see the effect of the local $S_N$-SSR on) correlations
between Alice's atoms and Bob's atom without considering entangled
states or mixed states. Here the states $\ket{\uparrow}$ and
$\ket{\downarrow}$ are orthogonal states of
the nucleus (spin up and spin down). 

Now if Alice's local operations (acting only on $A$s and $A'$s)
cannot distinguish molecules 1 and 2, then this state is {
equivalent} to \beq { \hat{T}_A(p_1)\ket{\psi} =
\ket{\downarrow_A^1
\downarrow_{A'}^1\uparrow_B^1}\ket{\uparrow_A^2 \uparrow_{A'}^2
\downarrow_B^2},} \eeq where $p_1$ is the swap permutation. Thus
under the action of ${\cal P}_A$ (or ${\cal P}_B$, or ${\cal
P}_A\otimes {\cal P}_B$), $\ket{\psi}$ goes to an { equal
mixture}: \bqa \ket{\psi} &\stackrel{{\cal P}_A \otimes {\cal
P}_B}{\longrightarrow} &
{\cal P}_A \otimes {\cal P}_B[\ket{\psi}\bra{\psi}] \nonumber\\
&=&  \uplus \smallfrac{1}{\sqrt{2}} \ket{\uparrow_A^1
\uparrow_{A'}^1 \uparrow_B^1}\ket{\downarrow_A^2
\downarrow_{A'}^2\downarrow_B^2} \nl{\uplus}
\smallfrac{1}{\sqrt{2}} \ket{\downarrow_A^1
\downarrow_{A'}^1\uparrow_B^1}\ket{\uparrow_A^2 \uparrow_{A'}^2
\downarrow_B^2}\label{Mixture} \eqa \blk Recall the notation
$\uplus$ defined in Sec.~II as a shorthand for describing a
projector. The two terms in the mixture are due to the two
elements in the $S_2$ group. Thus, under the $S_N$-SSR Alice knows
that both her atoms' spins are aligned. However, she loses
knowledge of their orientation with respect to Bob's atom.
Similarly, applying the $S_N$-SSR locally for Bob causes him to
lose information about the orientation of his spin with respect to
Alice's atoms.

\subsection{General Action of ${\cal P}$}\label{HilSpace}

Consider the general action of ${\cal P}$ on $N$ copies of a
$d$-dimensional system.  For our purposes $d$ is the total
Hilbert space dimension of a single molecule 
in the ensemble.  For example, if the molecules are made up of $M$
qubits, then $d=2^M$.  The general action of ${\cal P}$ can be
understood by analyzing the structure that it induces on the
Hilbert space of the total system, $(\mathbb{C}_d)^{\otimes N}$.
When the $S_N$-SSR applies to the system, as is the case for an
ensemble of identical particles or subsystems, this Hilbert space
carries a reducible representation $\hat{T}$ of $S_N$. Recall from
Sec. \ref{Technicalities} that this splits the Hilbert space into
`charge sectors':
\begin{equation}\label{SNDecomposition}
    (\mathbb{C}_d)^{\otimes N} = \bigoplus_{y \in Y} \mathbb{C}_y
    \,.
\end{equation}
The sectors are further decomposed into irreps of $S_N$:
\begin{equation}\label{SNIrepDecomposition}
    \mathbb{C}_y =  \mathbb{M}_y \otimes \mathbb{Q}_y
    \,,
\end{equation}
where $\mathbb{M}_y$ carries an irrep $\hat{T}_y$ of $S_N$,
$\mathbb{Q}_y$ carries the trivial irrep and has dimension given
by the multiplicity of $\hat{T}_y$ in $\hat{T}$. The label $y$ can
now be interpreted as a Young frame corresponding to an irrep of
$S_N$. The set of Young frames $Y$, viewed as Young diagrams, are
those consisting of $N$ boxes in up to $d$ rows of non-increasing
length. We define $D_y \equiv \text{dim}(\mathbb{M}_y)$. For
further details on the representations of $S_N$, see~\cite{Ful91}.

\subsubsection{Spin-1/2 particles}\label{spinhalf}
There are two cases where the structure of the Hilbert space induced
by ${\cal P}$ is particularly straightforward.  The first is when
the subsystems are identical spin-$\half$ particles. This means that
the ensemble is composed of $d=2$ dimensional systems and the
possible Young diagrams are those consisting of $N$ boxes in no more
than $d=2$ rows.  This limits the set of possible Young frames $Y$
to having $\lfloor N/2\rfloor+1$ elements, where $\lfloor
N/2\rfloor$ is the largest integer less than or equal to $N/2$. Thus
we are able to label each element by a single number. In this case,
since we are dealing with spin systems, it is sensible to set the
label $y$ for the Young frames equal to $j$, the ``total angular
momentum" of the ensemble.

Consider the one-party case of $N=2J$ spin-$\half$ particles (i.e.
$M=1$ qubit per molecule). The Hilbert space for each of the
particles is given by the $2$-dimensional complex vector space,
$\mathbb{C}_2$.  Using Eqs. (\ref{SNDecomposition}) and
(\ref{SNIrepDecomposition}) along with the fact that there are
$\lfloor J\rfloor+1$ Young frames labelled by $j$, the total Hilbert
space can be
decomposed into 
\begin{equation} \left(\mathbb{C}_2\right)^{\otimes 2J} =
\bigoplus_{j=J-\lfloor J\rfloor}^{J}\mathbb{M}_{j}
  \otimes \mathbb{Q}_{j}.\end{equation}
$\mathbb{M}_{j}$ and $\mathbb{Q}_{j}$ correspond to permutation
and angular momentum subspaces respectively. Thus permutations of
the spins $\hat{T}(p)$ act only upon $\mathbb{M}_{j}$ and joint
operations such as rotations act only upon $\mathbb{Q}_{j}$. The
dimensions of the subspaces are \begin{equation}{\rm
dim}(\mathbb{M}_{j})=d_j \equiv
\binom{2J}{J-j}\frac{2j+1}{J+j+1},\end{equation} for the
permutation subspace and\begin{equation}
 {\rm dim}(\mathbb{Q}_{j})=2j+1,\end{equation} for the angular momentum subspace.

Thus the basis for $\mathbb{C}_2^{\otimes 2J}$ in terms of these
subspaces can be written as:
$\cu{\ket{j,n}\otimes\ket{j,m}{\,:\,}_{j=J-\lfloor
J\rfloor}^{J}{\,;\,}_{m=-j}^{j}{\,;\,}_{n=1}^{d_j}}$, where $n$ is
a permutation label and $m$ is the magnetic quantum number. Now
consider the action of the permutation operator ${\cal P}$.
Physically, this operator destroys coherence between the `charge
sectors' and also acts to randomly permute the particles.
Mathematically this corresponds to $\mathcal{P}$ having the
following effect on a state matrix $\rho$ for an ensemble of
$N=2J$ qubits,
\begin{equation}\mathcal{P}\rho=\sum_{j=J-\lfloor J\rfloor}^{J}\mathcal{D}_j \otimes
\mathcal{I}_j(\hat\Pi_j \rho \hat\Pi_j ).
\label{EGspinhalf}\end{equation} Here $\hat{\Pi}_j$ is the
projection onto the charge sector $j$, and $\mathcal{D}_j$ is the
trace-preserving map that acts on the permutation subspace to
completely mix over the $\ket{j,n}$ basis states. $\mathcal{I}_j$
is the identity map over operators for the angular momentum
subspace, $\mathbb{Q}_j$.

\subsubsection{Ensemble of two molecules}\label{N2}
The second instance where it is straightforward to study the Hilbert
space structure is when there are only two molecules in the ensemble
(that is, $N=2$, so the $S_2$ group applies). In general the
molecules are $d$-dimensional systems, so the Hilbert space for each
molecule is given by $\mathbb{C}_d$. In this case there are only two
possible Young frames, corresponding to the symmetric and
antisymmetric representations of $S_2$.  These are both
$1$-dimensional representations meaning that the total Hilbert space
can be decomposed as,
\begin{eqnarray} \left(\mathbb{C}_d\right)^{\otimes 2}& =& \bigoplus_{y=s,a}
\mathbb{M}_1
\otimes \mathbb{Q}_y \nonumber\\
& =& \bigoplus_{y=s,a} \mathbb{Q}_y,\end{eqnarray} since $D_1={\rm
dim}(\mathbb{M}_1)=1$. The components of the angular momentum
subspace, $\mathbb{Q}_s$ and $\mathbb{Q}_a$, correspond to
symmetric and antisymmetric subspaces respectively. Their
dimensions are given by dim$(\mathbb{Q}_s)=(d^2+d)/2$ and
dim$(\mathbb{Q}_a)=(d^2-d)/2$.

This structure can be simply understood from the fact that there
are only two permutations in the $S_2$ group, which can be
represented by $\hat{T}(p_0)=\hat{I}$, and the operator
$\hat{T}(p_1)=\hat{T}$ which swaps the two molecules. The group
structure of $S_2$ ensures that $\hat{T}^2=\hat{I}$, which means
that $\hat{T}$ can be written as
$\hat{T}=\hat{\Pi}_s-\hat{\Pi}_a$, where the operators
$\hat{\Pi}_s$ and $\hat{\Pi}_a$ project onto the symmetric and
antisymmetric subspaces respectively. Also note that the identity
operator can be represented as $\hat{I}=\hat{\Pi}_s+\hat{\Pi}_a$.
Hence the action of ${\cal P}$ on the density matrix $\rho$ for an
$N=2$ ensemble state is given by
\begin{equation}
\mathcal{P}\rho=\frac{1}{2}\left(\hat{I}\rho\hat{I}+\hat{T}\rho\hat{T}^\dagger\right).
\label{N2ensemble}\end{equation} Using the expressions for
$\hat{I}$ and $\hat{T}$ in terms of projection operators gives,
\begin{eqnarray}
\mathcal{P}\rho&=&\frac{1}{2}[(\hat{\Pi}_s+\hat{\Pi}_a)\rho(\hat{\Pi}_s+\hat{\Pi}_a)+(\hat{\Pi}_s-\hat{\Pi}_a)\rho(\hat{\Pi}_s-\hat{\Pi}_a)]\nonumber\\
&=& \hat{\Pi}_s\rho\hat{\Pi}_s+\hat{\Pi}_a\rho\hat{\Pi}_a.
\label{N2ensembleProj}\end{eqnarray}  This illustrates the fact
that $\mathcal{P}$ destroys coherence between the angular momentum
`charge sectors', which in this case means destroying coherence
between the symmetric and antisymmetric subspaces.

\subsection{Multiple Copies under the
$S_N$-SSR}\label{sec:SSR:mcopies}
We have seen in Sec. \ref{sec:EntSSR} that pure states subject to a
SSR show remarkable similarities to mixed states.  In order to
obtain entanglement from mixed states we often consider preparing
many copies of the state and performing distillation protocols to
recover maximally entangled states. Similarly for pure states
subject to a SSR, it is possible to use many copies of the state to
obtain extractable entanglement.

\begin{figure}
\begin{center}
\epsfig{file=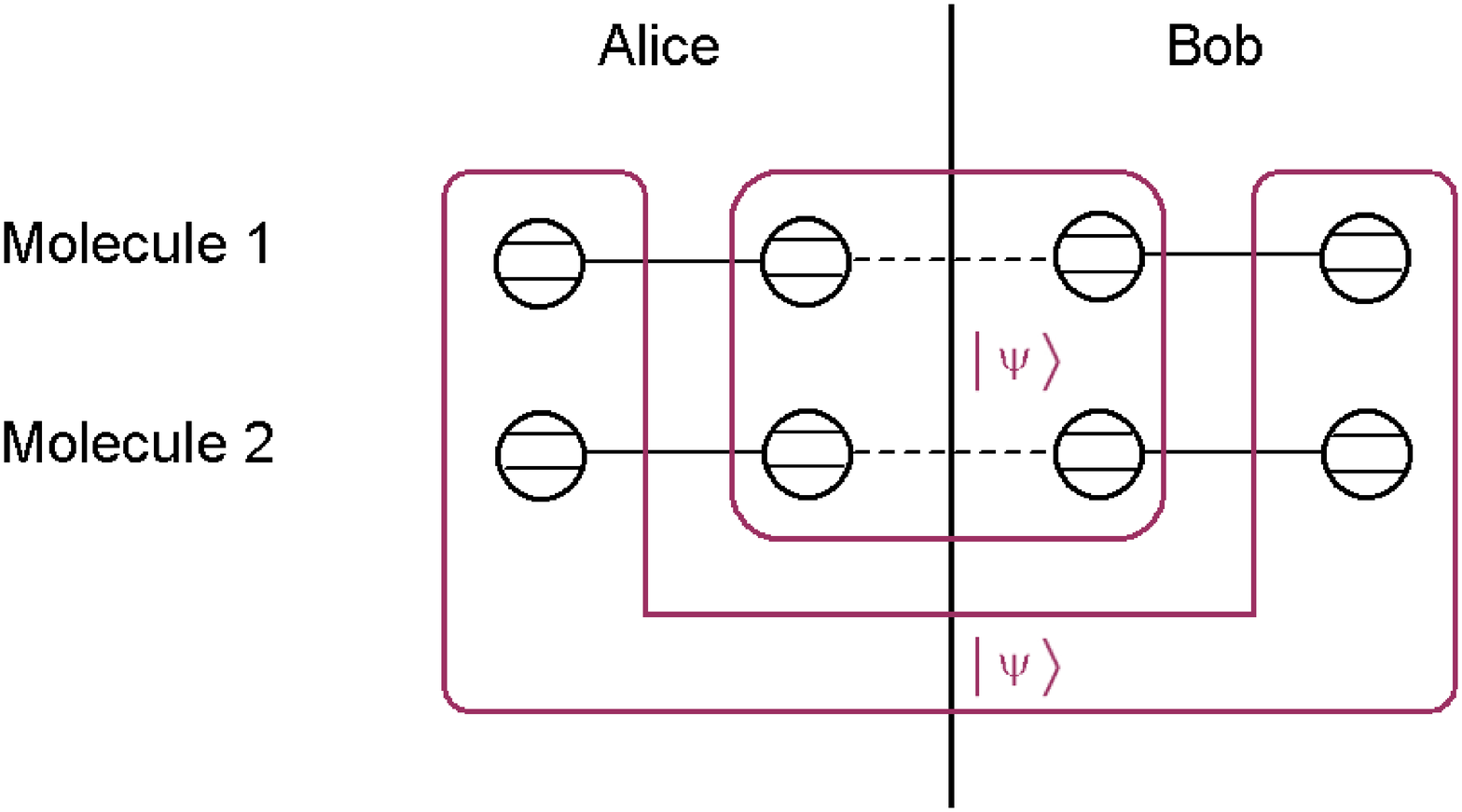, height=4.0cm,width=6.5cm}
\end{center}
\caption{\blk Creating multiple copies of an ensemble described by
$\ket{\psi}$. $N=2$ and $M'=4$, which means that Alice and Bob
share two copies of the $N=2$, $M=2$ ensemble.}\label{Multicopies}
\end{figure}

However, care must be taken when applying the notion of multiple
copies to ensemble states which are subject to the $S_N$-SSR. If
one were simply to double the number of molecules in the ensemble,
there would be more possible ways of permuting them and the system
would in fact be constrained by a different SSR (i.e. $S_{2N}$-SSR
rather than $S_N$-SSR). Applying the notion of multiple copies
under the $S_N$-SSR means duplicating an ensemble of $N$
molecules, each with $M$ atoms, by creating an ensemble of $N$
molecules, each with $2M$ atoms.  This way, each molecule now
contains two copies of the original state, and Alice and Bob
possess two copies of the original ensemble. In general they can
obtain $C$ copies of the original ensemble by increasing the
number of atoms in each of the $N$ molecules to $M'=CM$. If the
original ensemble of $N=2$ molecules had $M=2$ atoms (with Alice
and Bob each `owning' one atom from each molecule in the original
ensemble), two copies of the ensemble is given by an ensemble of
$N=2$ molecules with $M'=4$ atoms. This is illustrated in Fig.
\ref{Multicopies}. This concept will be discussed further in the
context of recovering entanglement ostensibly lost due to the SSR.

\section{Applications of $S_N$-SSR}

\subsection{ Asymptotic Loss of Entanglement}\label{Loss}


Typically with the $S_N$-SSR applied to many identical copies of
an entangled state the amount of extractable entanglement will be
less than in the unconstrained case. This was actually done first
by Eisert \ea\ \cite{Eis00}. Consider an ensemble of $N=2J$
identically prepared molecules each consisting of two nuclei in
the following state:
\begin{equation}
\ket{\psi} = \alpha\ket{\downarrow_A \downarrow_B} +
\beta\ket{\uparrow_A \uparrow_B},\label{spinstate}
\end{equation}
where $\alpha$ and $\beta$ can be taken to be real, so that
$\alpha^2+\beta^2=1$ is the normalization condition. Using the
Hilbert space decomposition into permutation and angular momentum
subspaces from Sec. \ref{spinhalf}, the total state of the ensemble
can be written as, \bqa \ket{\psi}^{\otimes N} &=& \sum_{j=J-\lfloor
J\rfloor}^{J}\sum_{n=1}^{d_j}
\sum_{m=-j}^{j} \alpha^{J-m}\beta^{J+m} \nonumber \\
&& \times \ket{j,n}_A \ket{j,m}_A\, \otimes\,
\ket{j,n}_B\ket{j,m}_B, \label{BasisState}\eqa where the condition
$\alpha^2+\beta^2=1$ also indicates that $\ket{\psi}^{\otimes N}$ is
normalized. From the spin representation [Eq. (\ref{spinstate})] it
is easy to see that the entanglement for the ensemble is
$E(\ket{\psi}^{\otimes N}) = N (-\alpha^2\log\alpha^2 - \beta^2
\log\beta^2)= N E(|\psi\rangle)$.

We now consider the amount of extractable entanglement under the
{$S_N$-SSR}. To do so, we must take into account the effect of the
SSR on the ensemble state. The permutation operator ${\cal P}$
results in a completely mixed state for both Alice and Bob in the
permutation subspace. That is, \bqa \ket{\psi}^{\otimes N}\!\!\!
&&\stackrel{{\cal P}_A \otimes {\cal P}_B}{\longrightarrow} \ \
\sum_{j=J-\lfloor J\rfloor}^{J}\; \sum_{n=1}^{d_j}\
\frac{I_A^j}{d_j}\otimes\frac{I_B^j}{d_j}\nonumber
\\ &\otimes&\!\!\left(\uplus\sum_{m=-j}^{j}
\alpha^{J-m}\beta^{J+m}\ket{j,m}_A \ket{j,m}_B\right)\!\!,
\label{43}\eqa where implementation of the $S_N$-SSR has also
destroyed coherence between different $j$ terms.  Note that
$I_A^j$ is the identity operator on the Hilbert space
$\mathbb{M}_j$ for Alice, and similarly $I_B^j$ for Bob. Eq.
(\ref{43}) can be simplified by defining the (normalized) angular
momentum part of the state as $\ket{\phi_j} =(1/\sqrt{d_j \wp_j}\
) \sum_{m=-j}^{j} \alpha^{J-m}\beta^{J+m} \ket{j,m}_A \otimes
\ket{j,m}_B$. The term $\wp_j = \sum_{m=-j}^j
\alpha^{2(J-m)}\beta^{2(J+m)}/d_j$ is the probability of obtaining
the $j$th angular momentum value and a particular irrep, indexed
by $n_A$ and $n_B$. Since there are actually $d_j^2$ irreps for
each $j$ value, the probability of obtaining a particular $j$ is
$d_j^2\wp_j$. It can be verified that $\sum_{j=J-\lfloor
J\rfloor}^J d_j^2\wp_j=1$ as required by conservation of
probability. Using these definitions allows \erf{43} to be
rearranged as,
\begin{eqnarray}
{\cal P}_A\!&\otimes&\!{\cal P}_B\ \left(\ket{\psi}\bra{\psi}\right)^{\otimes N} \nonumber \\
&=&\sum_{j=J-\lfloor J\rfloor}^{J}\; \!\!\!\!\!d_j
\left(\frac{I_A}{d_j}\otimes\frac{I_B}{d_j}\right) \otimes\
d_j\wp_j\left(\frac{}{}\!\!\uplus\ket{\phi_j}\right)\!.
\end{eqnarray}

For convenience we will omit writing the completely mixed states
on the permutation subspace, although when we write the
$S_N$-invariant state they are assumed to be there. Using this
convention, the $S_N$-invariant state can be written compactly as
\begin{equation}
{\cal P}_A\otimes{\cal
P}_B\left(\ket{\psi}\bra{\psi}\right)^{\otimes N} =\sum_{j=J-\lfloor
J\rfloor}^{J}\;d_j^2\wp_j\ \left(\!\!\frac{}{}
\uplus\ket{\phi_j}\right)\!.\label{SNcompact}
\end{equation}

Since no observed quantities can be changed by replacing
$\ket{\psi}^{\otimes N}$ with the $S_N$-invariant state,
calculating the constrained entanglement of $\ket{\psi}^{\otimes
N}$ is equivalent to
\begin{equation} E_{S_N\textrm{-SSR}}(\ket{\psi}^{\otimes N}) = E_D \ro{ {\cal
P}_A\otimes {\cal P}_B \left(\ket{\psi}\bra{\psi}\right)^{\otimes
N} }.\label{ConstrainedE}\end{equation} If the state of interest
is composed of states that are are locally distinguishable (for
both Alice and Bob) then it is known as a \emph{biorthogonal
mixture} (see Ref. \cite{Herbut03} for more details). The expected
entanglement of such a state is simply a weighted sum of the
entanglement present in each of the the possible states. That is,
\begin{equation} E(\rho)=\wp_1E(\rho_1)+\wp_2E(\rho_2)+\ldots,\label{ExpectedE}\end{equation}
where $E(\rho_{1,2,\ldots})$ is the entanglement of the locally
distinguishable states making up the mixture, and $\wp_{1,2,\ldots}$
are the corresponding probabilities of each state occurring. The
$S_N$-invariant state is of this form, with $\uplus\ket{\phi_j}$ the
possible states and $d_j^2\wp_j$ the corresponding probabilities.
Therefore, Eq. (\ref{ConstrainedE}) can be rewritten as
\begin{equation} E_{S_N\textrm{-SSR}}(\ket{\psi}^{\otimes N}) =
\sum_{j=J-\lfloor J\rfloor}^{J} d_j^2 \wp_j E(\ket{\phi_j}),
\end{equation} where $E(\ket{\phi_j})$ is the entanglement of the
angular momentum state $\ket{\phi_j}$. We expect the total amount of
constrained entanglement to be less than the $E(\ket{\psi}^{\otimes
N}) = N (-\alpha^2\log\alpha^2 - \beta^2 \log\beta^2)$ ebits
calculated for the unconstrained system.

To demonstrate this, consider the particular case of Bell states,
where $\alpha=\beta=\smallfrac{1}{\sqrt{2}}$. This gives $
E(\ket{\psi}^{\otimes N}) = N$, but, as shown by Bartlett and
Wiseman \cite{BarWis03}, \bqa
  E_{S_N\textrm{-SSR}} ( \ket{\psi}^{\otimes N} ) &=&\!\!\!\!
\sum_{j=J-\lfloor J\rfloor}^{J}\!\! d_j^2 \wp_j\log_2 (2j+1).\label{Eloss}  
\eqa This expression can be simplified significantly in the
asymptotic limit (i.e. $J=N/2\rightarrow \infty$) because the
probability distribution $ d_j^2 \wp_j$ becomes sharply peaked at
a single $j$ value. Thus, a single term in the sum essentially
determines the value of the entanglement. It can be shown that for
large ensembles ($N\gg1$) the significant term in the sum is
specified by $j\approx \sqrt{J}$. This means that in the
asymptotic limit Eq. (\ref{Eloss}) reduces to approximately $(1/2)
\log_2 N$. Since this is the maximum total entanglement, the
entanglement per molecule must always $\to 0$ as $N \to \infty$.
Hence, under the $S_N$-SSR for an ensemble of maximally entangled
pure states we asymptotically lose the ability to access the
entanglement.

\subsection{Asymptotic Recovery of Entanglement}\label{Recovery}

We have just shown that under the $S_N$-SSR we apparently `lose'
much of the entanglement in the ensemble. This might seem contrary
to the intuition obtained from the U(1) case, for example, where
in the limit of a large number of particles, the entanglement per
particle is \emph{recovered} asymptotically approaching the
unconstrained entanglement \cite{WisVac03}. This
discrepancy arises from taking an inappropriate form of the asymptotic
limit for the $S_N$-SSR. As explained in Sec.
\ref{sec:SSR:mcopies}, having multiple copies under an $S_N$-SSR
does not mean changing $N$. The asymptotic limit for the number of
copies thus should be considered with $N$ fixed.

We begin by considering an ensemble of $N=2$ molecules. As discussed
in Sec. \ref{N2} this is a special case that considerably simplifies
the action of $\mathcal{P}$.  To relate to Sec. \ref{Loss}, imagine
that Alice and Bob share an ensemble of two molecules each of which
is a Bell state. The difference here is that we allow each molecule
to be larger and to contain $C$ copies of a Bell state. That is, we
allow Alice and Bob to share $C$ copies of the original $N=2$
ensemble.

For convenience we define the density matrix for $C=1$ copy of the
ensemble of $N=2$ Bell states as
\begin{equation} \rho_{AB}=\Big[\ket{\psi^-}\bra{\psi^-}\Big]^{\otimes
2},\end{equation} where the Bell singlet state \footnote{For
simplicity with our formalism we make use of the singlet Bell
state, however, our results also hold for the triplet Bell
states.} is defined as
$\ket{\psi^-}=\frac{1}{\sqrt{2}}(\ket{\uparrow_A\downarrow_B}-\ket{\downarrow_A\uparrow_B})$.
The state $\rho_{AB}$ can also be expressed as
\begin{eqnarray}
\rho_{AB}&=&\uplus\frac{1}{2}\Big[\ket{A}+\sqrt{3}\ket{S}\Big],
\end{eqnarray} where we define normalised states in the antisymmetric and symmetric
subspaces in terms of the $\ket{j,m}$ basis (recall Sec.
\ref{spinhalf}) as $\ket{A}=\ket{j=0,m=0}_A\ket{j=0,m=0}_B$ and
$\ket{S}=(1/\sqrt{3})\sum_{m=-1}^1\ket{j=1,m}_A\ket{j=1,-m}_B$
respectively.

Using this representation for the state, it becomes apparent that
$\mathcal{P}$ simply destroys coherence between the symmetric and
antisymmetric subspaces which can be represented as
\begin{equation}\mathcal{P}\rho_{AB}=\uplus\sqrt{\frac{1}{4}}
\ket{A}\uplus\sqrt{\frac{3}{4}}\ket{S}.
\label{Erecover}\end{equation}

Since Alice and Bob share a biorthogonal mixture, they can each
make local measurements to distinguish between the symmetric and
antisymmetric subspaces. This is equivalent to the situation
considered by Eisert \emph{et al.} \cite{Eis00}. With probability
$1/4$ they find that they have the locally antisymmetric state and
they retain no entanglement (as this is a separable state).
However, with probability $3/4$ they obtain the locally symmetric
state, which is equivalent to a maximally entangled qutrit state.
In that case they retain $\frac{3}{4}\log_2(3)\approx 1.19$ ebits
of entanglement. Without the $S_2$-SSR constraining their two Bell
states, Alice and Bob would possess 2 ebits of entanglement.

\begin{figure}
\begin{center}
\epsfig{file=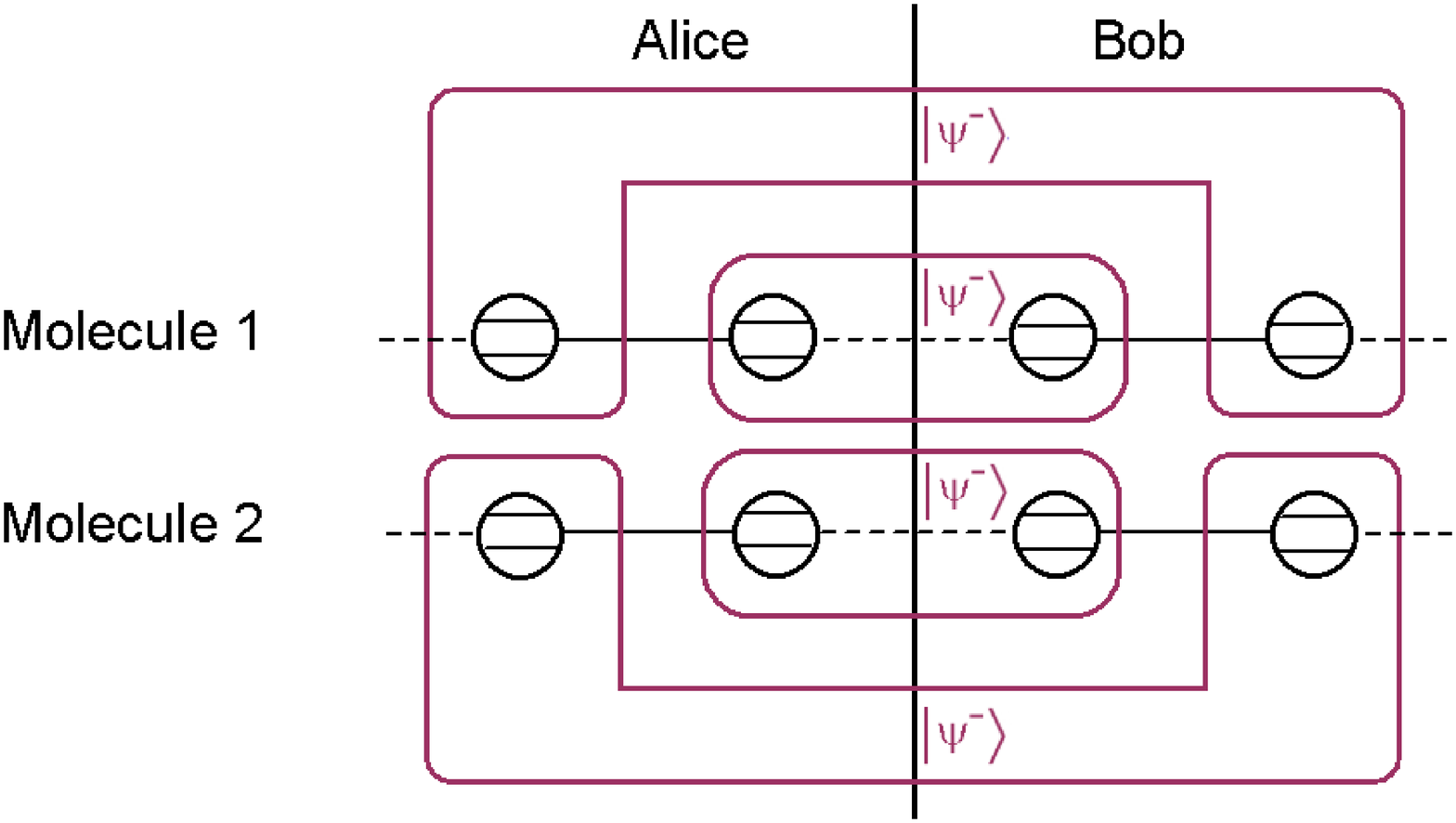, height=4.0cm,width=6.5cm}
\end{center}
\caption{\blk Two copies ($C=2$) of $\rho_{AB}$ (which is composed
of two Bell states $\ket{\psi^-}$). Each molecule can be extended
to include more Bell states to increase the number of copies $C$
of $\rho_{AB}$.}\label{BellStates}
\end{figure}

One might expect that by using the concept of multiple copies it
would be possible to ameliorate the effect of the SSR. This is
indeed the case, as we now show. For the $S_2$-SSR to apply, Alice
and Bob must share entanglement contained in 2 molecules.  In the
simplest case, each molecule is simply a Bell singlet state and
the combined state is $\rho_{AB}$, as discussed above. To apply
the concept of multiple copies, Alice and Bob must share $C$
copies of $\rho_{AB}$ (see Fig. \ref{BellStates}). With no
restrictions in place Alice and Bob would share $2C$ ebits of
entanglement.

The calculation of how much entanglement is retained using
multiple copies can be significantly simplified by noting that in
this case, each of the molecules (containing $C$ Bell pairs) can
be considered as a maximally entangled \emph{qudit} pair. This is
possible due to the global symmetry of the ensemble state chosen.
In this case, each molecule can be described as a maximally
entangled pair of qudits, with the qudits dimension given by
$d=2^C$. This simplifies calculations, as the maximum entanglement
of a pair of entangled qudits is readily calculated to be $E_{\rm
max}=\log_2 d$. Thus, without considering the $S_2$-SSR
constraint, the total entanglement for the two maximally entangled
qudit pairs is $E=2C$ ebits, as already derived.

We can express the state of $C$ copies of $\rho_{AB}$ under the
$S_2$-SSR explicitly as a biorthogonal mixture of a locally
symmetric and a locally antisymmetric state,
\begin{equation}\mathcal{P}\left(\rho_{AB}\right)^{\otimes C}=\wp_s\rho_s+\wp_a\rho_a,\label{biomix}\end{equation}
where the weightings $\wp_s$ and $\wp_a$ are the probabilities of
both Alice and Bob obtaining a locally symmetric or locally
antisymmetric state respectively. These probabilities depend upon
the dimension of the subspace that each of the local states
occupy: $\wp_s={\rm dim}\left(\mathbb{Q}_s\right)/d^2$ and
$\wp_a={\rm dim}\left(\mathbb{Q}_a\right)/d^2$ (recall the
expressions for the subspace dimensions defined in Sec. \ref{N2}).

The structure of Eq. (\ref{biomix}) means that it is quite
straightforward to calculate the extractable entanglement of
$\left(\rho_{AB}\right)^{\otimes\ C}$. It is simply a weighted
average of the entanglement in the two subspaces:
\begin{equation}E=\frac{d^2-d}{2d^2}\log_2\!\!\left(\frac{d^2-d}{2}\right)+\frac{d^2+d}{2d^2}\log_2\!\!\left(\frac{d^2+d}{2}\right).\label{Ent}\end{equation}
For a large number of copies ($C \gg 1$ ) the dimension $d$ is
large and Eq. (\ref{Ent}) reduces to approximately $E = 2C-1$.
Thus in the asymptotic limit, nearly all of the entanglement has
been recovered (only a single ebit has been lost).

\begin{figure}
\begin{center}
\epsfig{file=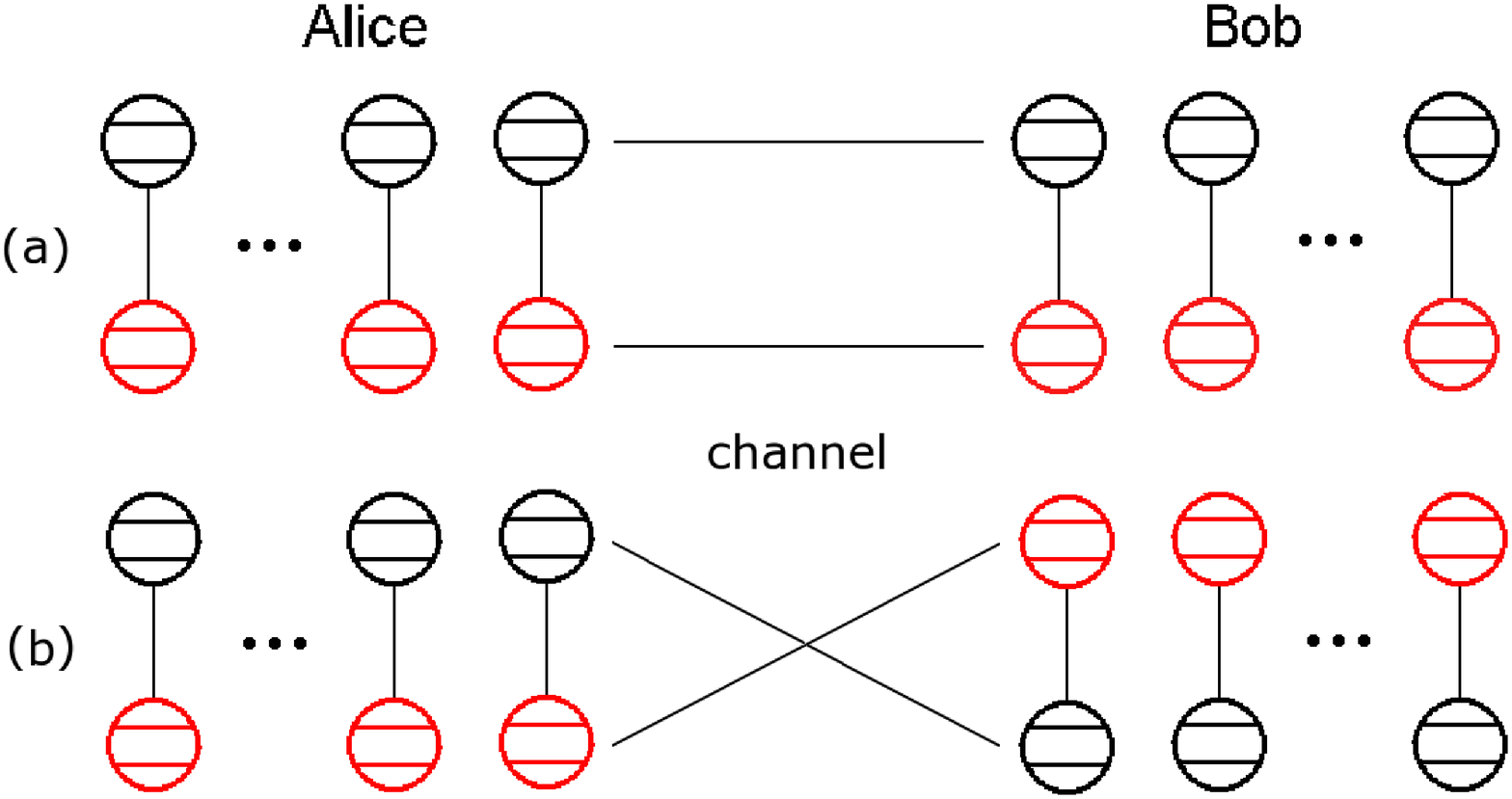, height=4.0cm,width=6.5cm}
\end{center}
\caption{\blk Alice and Bob share $N$ copies of $\rho_{AB}$ via a
channel.  In case (a) the channel distributes the states in order.
In (b) the channel swaps the ordering within each
pair.}\label{Channel}
\end{figure}

Another way to consider this problem is that Alice and Bob share
many copies of the state $\rho_{AB}$ via a channel (see Fig.
\ref{Channel}). The channel is deterministic and either does
nothing or performs a swap of the molecules. If Alice and Bob were
unable to make collective measurements on their entire collection
of qubits then they could still make use of their copies of
$\rho_{AB}$ to asymptotically retain much of their entanglement. A
non-optimal procedure that they could implement would be to use up
a small number of copies to find out what map the channel performs
(either identity or swapping). Once they know what the channel
does they can then safely use the 1 ebit of entanglement in each
of their remaining Bell pairs. This method is non-optimal because
Alice and Bob lose at least a few ebits of entanglement in
characterizing the channel (and asymptotically with collective
measurements they need lose only 1 ebit).

In general for the case of the $S_N$-SSR with $N>2$ it is difficult
to optimally calculate the exact asymptotic amount of entanglement
recovered.  However, considering the non-optimal procedure just
discussed it is intuitive that Alice and Bob could recover most of
their entanglement (in the asymptotic limit) simply by using up some
copies of the state to characterize the `channel'. They would then
retain the entanglement in the remaining copies. As the size of the
ensemble ($N$) increases, more copies of the state will be required
to satisfactorily characterize the `channel' and thus more
entanglement will be lost.

\section{Reference Frames}

In general, a reference frame for a SSR is something that removes
its effect. For example, a perfect reference frame completely
removes the effect of, or `lifts', the SSR. This is the ideal
case, although in practice it is possible to have partial
reference frames which only partially remove the effect of the
SSR.

Usually a reference frame is an extra system added to the system
of interest which allows access to degrees of freedom otherwise
unaccessible due to the SSR. Thus, for an ensemble of molecules,
for which the $S_N$-SSR applies, one might naively expect to add
an extra ensemble of molecules to act as a reference frame.
However, as discussed in Sec. \ref{sec:SSR:mcopies}, due to the
nature of the $S_N$ group, adding molecules would in fact alter
the SSR for the system. That is, the reference molecules would
actually be permuted with the system molecules, making it more,
not less, difficult to gain information about the system.

Instead, the type of reference frame needed for an ensemble system
is analogous to a labelling.  Classically, one would think of
physically writing a label (say a number) on each object, to serve
as a reference ordering. Physically this corresponds not to adding
molecules to the ensemble, but adding an extra nucleus (or group
of nuclei) to each molecule in the ensemble.

\begin{figure}
\begin{center}
\epsfig{file=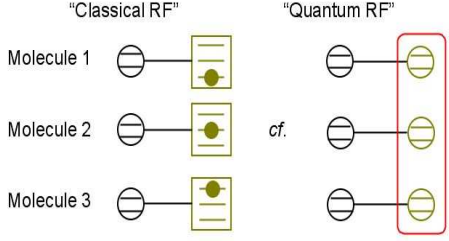, height=4.0cm,width=6.5cm}
\end{center}
\caption{\blk Classical versus quantum reference frames. The
classical reference frames on the left are represented by boxes
and are uncorrelated. On the right the reference frames are
quantum systems and we allow for correlations between the label
systems.}\label{ReferenceFrames}
\end{figure}

To illustrate this, consider a simple example, 
with $N=3$ molecules. In this instance a pure state [where the
three molecules happen to be uncorrelated, see Fig.
(\ref{ReferenceFrames})] with a reference frame is
\begin{equation}
\ket{\Psi} = \ket{\psi^1,1}\ket{\psi^2,2}\ket{\psi^3,3} =
\ket{\psi^1,\psi^2,\psi^3} \otimes { \ket{1,2,3}}
\end{equation}
Here $\ket{\psi^k}$ is the state of the $M$ nuclei in the $k$th
molecule (not including the reference frame) which we have assumed
to factorize. In regards to the tensor product structure it is
important to remember that the second system is not in the same
state as the first (it need not even have the same Hilbert space
dimension).

\subsection{Quantum reference frames}

In the classical example above, we placed each $N$-dimensional
attached label system (nucleus or group of nuclei) in a unique
product state.  An obvious question is whether or not it is possible
to use label systems of smaller dimension if we allow entanglement
between the states of the $N$ label systems.  As demonstrated by von
Korff and Kempe~\cite{vKK04}, it is indeed possible to reduce the
dimension of the label systems by a constant factor in the limit
$N\rightarrow \infty$.

Recalling the structure of the Hilbert space from Sec.
\ref{HilSpace}, a state of the $N$ label systems $|0\rangle \in
(\mathbb{C}_d)^{\otimes N}$ that works as a \emph{perfect} quantum
reference frame would satisfy the property that the $N!$ states
\begin{equation}\label{OrthogonalStates}
    |p_n\rangle = \hat{T}(p_n)|p_0\rangle \,,
\end{equation}
for all $p_n \in S_N$ satisfy $|\langle p_n| p_{n'} \rangle|^2 =
\delta_{n,n'}$. This property ensures that every different ordering
is classically distinguishable (i.e., is associated with an
orthogonal quantum state).  So the problem reduces to the following:
What is the minimum $d$ such that such a set of orthogonal states
exists?

First, we note that the space $\mathbb{H}_R$ spanned by $\{
|p_n\rangle,\, p_n\in S_N\}$ is $N!$-dimensional and that the
representation $\hat{T}$ when restricted to this space is
isomorphic to the (left) \emph{regular representation}.  (The
regular representation $R$ of a group $G$ has $G$ as a carrier
space, and acts as $R(g)g' = gg'$.)  It is
well-known~(\cite{Ful91}, p.~17) that the regular representation
of $S_N$ contains every irrep $\hat{T}_y$ of $S_N$, each with a
multiplicity equal to $D_y$, the dimension of $\hat{T}_y$. Thus,
for $\hat{T}$ to contain the regular representation, it must
contain every irrep $\hat{T}_y$ of $S_N$ with a multiplicity of at
least $D_y$. In particular, this must hold true for the
fully-antisymmetric representation of $S_N$ (the irrep labeled by
a Young diagram consisting of a single column of $N$ boxes), and
$\hat{T}$ only contains the fully-antisymmetric representation if
$d\geq N$. Thus, if we demand that the label systems act as a
\emph{perfect} reference frame for $S_N$, then each label system
must be at least $N$-dimensional.

However, von Korff and Kempe~\cite{vKK04} have shown that it is
possible to use label systems with any dimension $d > \lfloor N/e
\rfloor$ if the requirement of a perfect reference frame is
relaxed to the less-stringent demand that, for $p_n\neq p_{n'}$,
$|\langle p_n| p_{n'}\rangle|^2 \rightarrow 0$ as $N \rightarrow \infty$.
(That is, that the reference frame states are distinguishable only
in the asymptotic limit.)  The basic idea is that if $d > \lfloor
N/e \rfloor$ then, although $\hat{T}$ does not contain \emph{all}
irreps of $S_N$ with the required multiplicity, the set that are
missing has measure approaching zero as $N \rightarrow \infty$. We
refer the reader to~\cite{vKK04} for details.

We now explicitly construct states of the form of
Eq.~(\ref{OrthogonalStates}), using the general construction
of~\cite{KMP04} that was subsequently applied specifically to the
$S_N$ group in~\cite{vKK04}.  Let $\overline{Y}$ be the set of
irreps that are contained in $\hat{T}$ and have sufficient
multiplicity, i.e., that satisfy \blk $\text{dim}\mathbb{Q}_y \geq
D_y$.  For each $y \in \overline{Y}$, choose an arbitrary subspace
$\mathbb{Q}'_y \subset \mathbb{Q}_y$ of dimension $D_y$. Let $\{
|y,i,j\rangle, i,j=1,\ldots,D_y \}$ be a basis for $\mathbb{M}_y
\otimes \mathbb{Q}'_y$, where $i$ labels a basis for $\mathbb{M}_y$
and $j$ labels a basis for $\mathbb{Q}'_y$. Define $D = \sum_y
D_y^2$. Then the state
\begin{equation}\label{PermutationRFstate}
    |p_0\rangle = \sum_{y \in \overline{Y}} \sum_{i=1}^{D_y}
    \sqrt{\frac{D_y}{D}}|y,i,i\rangle \,,
\end{equation}
can be used to define a set of states $\{|p_n\rangle =
\hat{T}(p_n)|p_0\rangle \}$ for $p_n\in S_N$ as in
Eq.~(\ref{OrthogonalStates}).  As demonstrated in~\cite{vKK04},
$\lim_{N\to\infty} D = N!$ and $\lim_{N\to\infty}|\langle p_n| p_{n'} \rangle|^2 =
\delta_{n,n'}$ provided that $d > \lfloor N/e
\rfloor$.

\subsection{Shared reference frames}
The simplest \emph{shared} reference frame is for Alice and Bob
each to have a reference frame. In general, if Alice and Bob share
$N$ tensor product states and both have a reference frame for each
state, then the total system can be described as \bqa\ket{\Psi}
&=& \bigotimes_{i=1}^N\ket{\psi^i_{AB},i_A,i_B}.\eqa  \blk For
example, this can be written out explicitly for the case when two
product states are shared, \bqa\ket{\Psi} &=&
\ket{\psi^1_{AB},1_A,1_B}\ket{\psi^2_{AB},2_A,2_B}\nn\\
&=&\ket{\Psi_{AB}}\ket{p_0}_A\ket{p_0}_B,\eqa \blk where in the second
line we have written the shared states first, followed by Alice
and Bob's reference frames. Note that we have rewritten Alice's reference state
$\ket{\cdots,1_A,\cdots}\ket{\cdots,2_A,\cdots}$ as the fiducial reference
state $\ket{\cdots}\ket{p_0}_A\ket{\cdots}_B$, and similarly for Bob's.

Although these states are separable, they cannot be prepared {\em
locally} by { ${\cal P}_A\otimes {\cal P}_B$-invariant} operations
from a { ${\cal P}_A\otimes {\cal P}_B$-invariant} state. Hence
they are bound entangled states which may become locally
preparable. (Recall the definitions in Sec. \ref{sec:EntSSR}.)
Note that such states are not globally ${\cal P}$-invariant.
However, using the final reference frame basis above we can write
a separable ${\cal P}$-invariant reference frame:
\begin{equation}{ \biguplus_{p_n\in S_N} \frac{1}{\sqrt{N!}}\ket{p_n}_A\ket{p_n}_B}.\end{equation}
This reference frame is an incoherent mixture of
reference states which is an example of a \emph{shared} reference
frame. The key point is that the same permutation is applied to
both Alice and Bob's reference states resulting in perfect
correlation between each of Alice and Bob's labels. That is, this
reference frames gives no indication of labels for individual
states, but indicates that Alice and Bob's particles are in the
same order. States of this form are mixed (separable) and hence
not part of the classification scheme of Sec. \ref{sec:EntSSR}.

Alternatively, a pure globally ${\cal P}$-invariant reference
frame can be constructed by considering non-separable states:
\begin{equation}{ \sum_{p_n\in S_N} \frac{1}{\sqrt{N!}}\ket{p_n}_A\ket{p_n}_B}.\end{equation} 
This state is a coherent superposition of reference states which are
perfectly correlated between Alice and Bob. Once again for an
explicit example we consider a reference frame for the $S_2$ group
\begin{eqnarray}\ket{\Psi}_{\rm RF}&=&\frac{1}{\sqrt{2}}\sum_{p_n\in
S_2}\ket{p_n}_A\ket{p_n}_B\nonumber\\&=&\frac{1}{\sqrt{2}}
\left[\ket{p_0}_A\ket{p_0}_B+\ket{p_1}_A\ket{p_1}_B\right],\label{ReferenceFrameStateS2}
\end{eqnarray} where $p_0$ is the identity permutation and $p_1$ is the swap permutation. In this case it can
be shown that the partial transpose of the state matrix $\rho_{\rm
RF}=\ket{\Psi}_{\rm RF}\bra{\Psi}_{\rm RF}$ is actually equal to
$\rho_{\rm RF}$. Thus it is a valid state matrix which means that
$\rho_{\rm RF}$ has a positive partial transpose \cite{Peres96}.
This shows that for the $S_2$ group, which is actually an Abelian
group, a shared reference state of the form of Eq.
(\ref{ReferenceFrameStateS2}) is become 1-distillable (this is because it contains no
entanglement under the $S_2$-SSR but becomes 1-distillable if
the SSR is lifted).

\section{Analogies with mixed-state entanglement}

\subsection{Activation}

Recall from section \ref{Distillable} that a general state $\rho$
is called 1-distillable if by LOCC Alice and Bob can,
with some probability, create from it a nonseparable two-qubit
state. Also recall that there are bound entangled states that
become 1-distillable when the two parties have their LOCC
supplemented by a shared PPT-channel. These states, as we have
mentioned in section \ref{Become}, are called become 1-distillable states.

Since $S_N$ is a finite group, reference frames for the $S_N$ group
can be finite (this is quite different to the case for Lie group
SSRs such as the U(1)-SSR). Moreover, the $S_N$ reference frames can
be used without being disturbed because they form an orthonormal
set. Thus under the $S_N$-SSR there is no distinction between
activation of a bound entangled state (by a bound entangled state which becomes locally preparable) and
lifting the $S_N$-SSR to make become 1-distillable states 1-distillable.

Activation of a bound entangled state can be seen in the following
example. If $N=2$ and $M=2$, (i.e Alice and Bob own one nucleus
per molecule), then the state \begin{equation}{ \sqrt{2}\ket{\psi}
= \ket{+}_A\ket{-}_B + \ket{-}_A\ket{+}_B},\end{equation} is bound
entangled that can become 1-distillable. Here $\ket{+} =
\ket{j=1,m=0}$ and $\ket{-}=\ket{j=0,m=0}$, so
${\hat{T}(p_1)\ket{\pm}=\pm \ket{\pm}}$. From this it is easy to
see that $\ket{\psi}$ is globally symmetric, but under the local
SSR, \begin{equation}{ \sqrt{2}\ket{\psi} \stackrel{{\cal P}_A
\otimes {\cal P}_B}{\longrightarrow} \;\uplus\; \ket{+}_A\ket{-}_B
\;{ \uplus}\; \ket{-}_A\ket{+}_B},\end{equation} which is clearly
separable. Hence, with the SSR the state has no distillable
entanglement.

\begin{figure}
\begin{center}
\epsfig{file=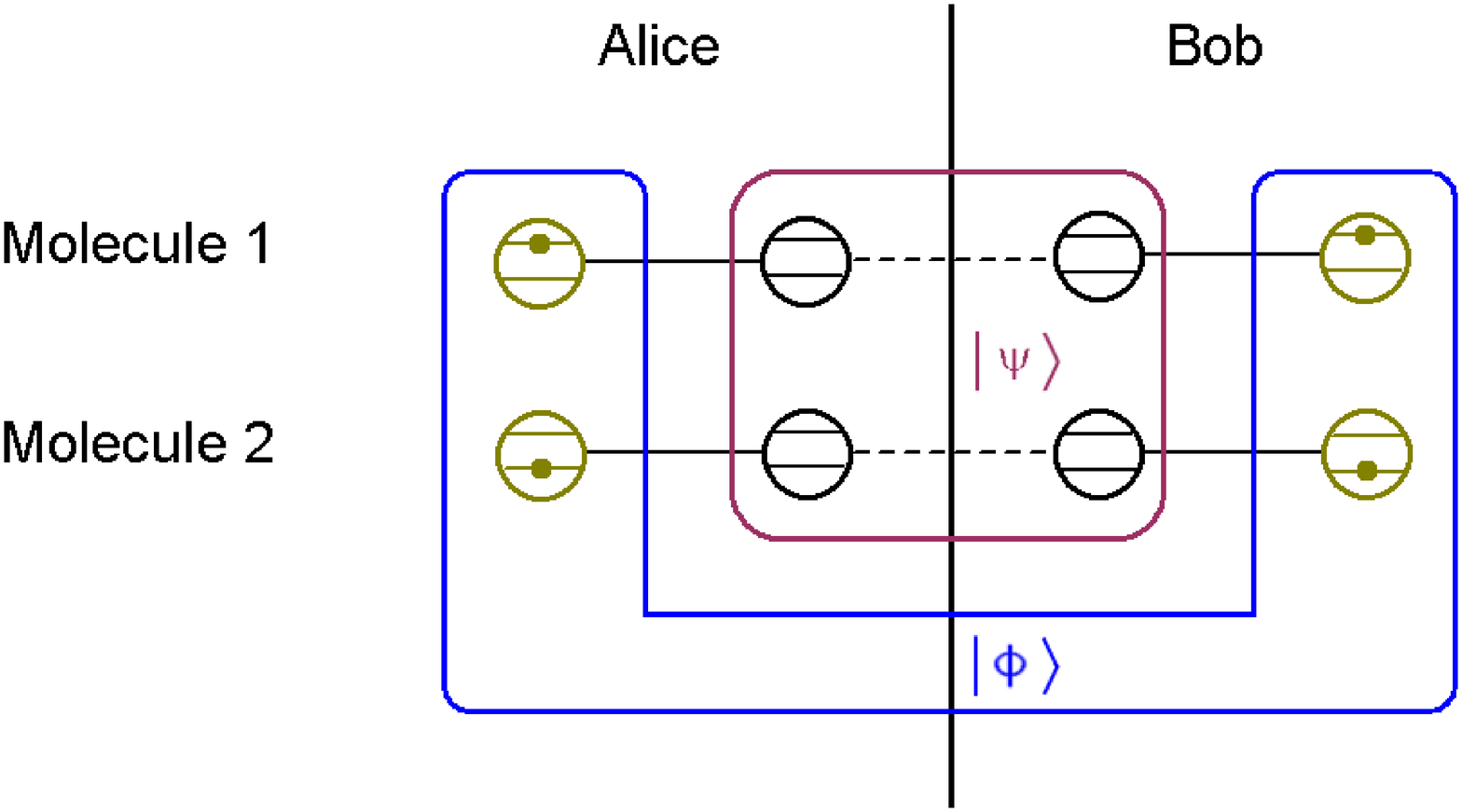, height=4.0cm,width=6.5cm}
\end{center}
\caption{\blk Using an extra state $\ket{\phi}$ to activate the
bound entanglement in $\ket{\psi}$. In this case $\ket{\phi}$ acts
as a perfect reference frame and all the entanglement in
$\ket{\psi}$ is recovered.}\label{Activate}
\end{figure}

It is possible to completely lift the SSR and regain 1-ebit of
entanglement from this state.  This is achieved by adding an extra
shared state $\ket{\phi}$ to \emph{activate} the bound
entanglement in $\ket{\psi}$. This is shown in Fig.
\ref{Activate}. For instance, the simplest perfect reference frame
$\ket{\phi}$ would label each of Alice and Bob's nuclei, for
example, $\ket{\phi}=\ket{1_A,2_A}\ket{1_B,2_B}=\ket{p_0}_A\ket{p_0}_B$.
Then it becomes possible for Alice to find out which of her nuclei
is correlated with which of Bob's simply through measurement of
the shared reference state. Thus by use of a reference frame (that
is, activating the bound entanglement), it is possible to access
1-ebit of entanglement from the become 1-distillable state.

\subsection{Distillation}

We now illustrate the phenomenon of distillation using the same
example state $\ket{\psi}$. That is, although without a reference
frame the state ${ \sqrt{2}\ket{\psi} = \ket{+}_A\ket{-}_B +
\ket{-}_A\ket{+}_B}$ has $E_{S_2\textrm{-SSR}}=0$, with two copies
some entanglement can be obtained.

\begin{figure}
\begin{center}
\epsfig{file=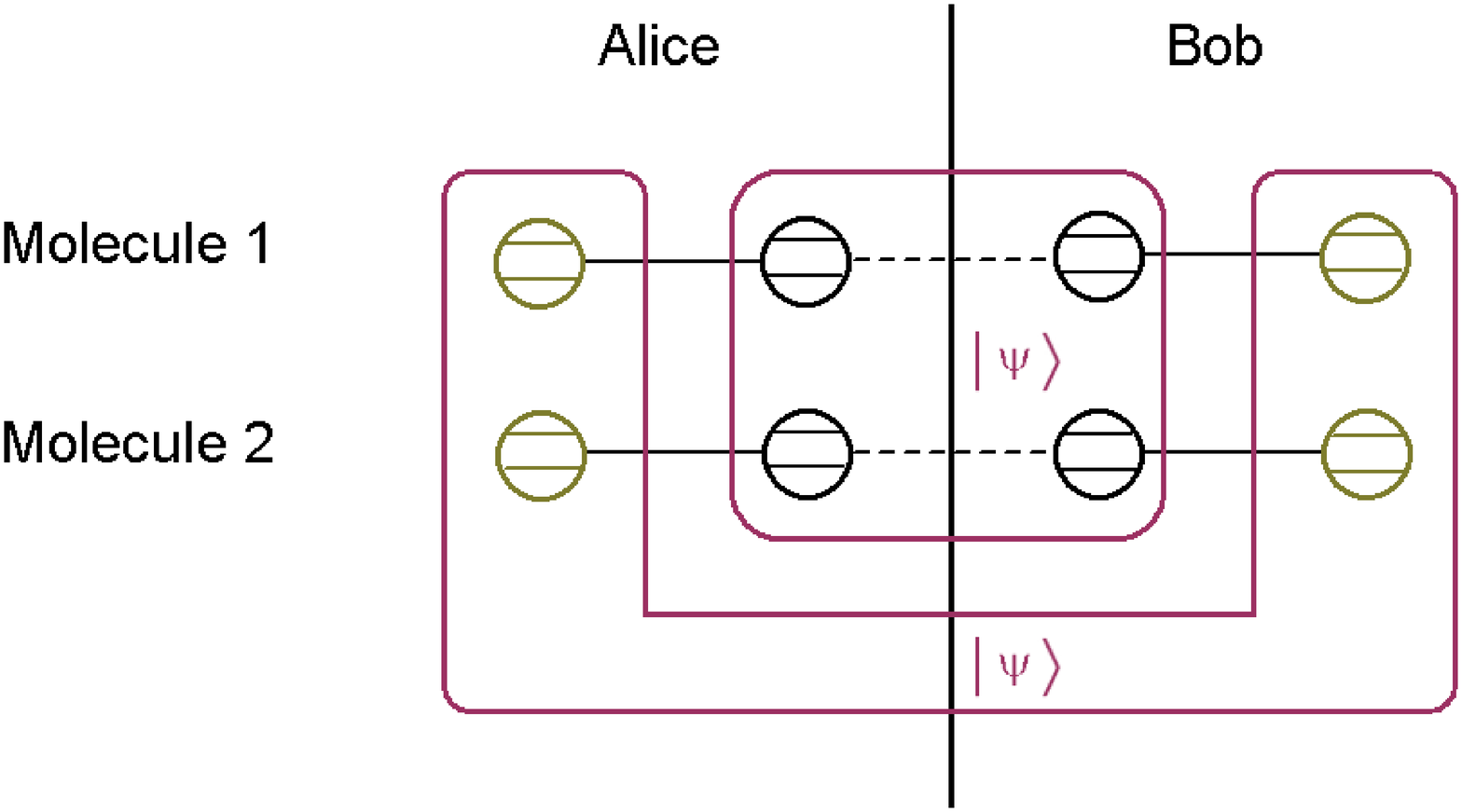, height=4.0cm,width=6.5cm}
\end{center}
\caption{\blk With two copies of the state $\ket{\psi}$ the second
can act as a reference frame for the first allowing one ebit of
entanglement to be accessed. This is considered an imperfect
reference frame for the system as we would expect two copies of
$\ket{\psi}$ to contain two ebits of entanglement.
}\label{Duplication}
\end{figure}

Recall from Section \ref{sec:SSR:mcopies} that two copies does
{\em not} mean four molecules. Since $S_2$ is fixed, we still have
$N=2$ molecules, but instead of $M=2$ we now have $M'=4$, that is,
Alice and Bob each have \emph{two} nuclei per molecule. This is
demonstrated in Fig. \ref{Duplication}.
The state for the two copies can be written as, \bqa
(\sqrt{2}\ket{\psi} )^{\otimes 2} &=& \ket{++}_A\ket{--}_B +
\ket{-+}_A\ket{+-}_B \nl{+} \ket{+-}_A\ket{-+}_B +
\ket{--}_A\ket{++}_B, \eqa which with a perfect reference frame
contains two ebits of entanglement. The effect of the SSR is to
create a mixture of the unchanged state with the state formed by
applying the swap $\hat{T}(p_1)$ to Alice's particles
(or Bob's). Thus under the $S_2$-SSR, the state becomes,
\bqa (\sqrt{2}\ket{\psi} )^{\otimes 2} &\stackrel{{\cal P}_A \otimes
{\cal P}_B}{\longrightarrow}& \uplus \left( \ket{++}_A\ket{--}_B +
\ket{--}_A\ket{++}_B\right) \nl{\uplus} \left(\ket{-+}_A\ket{+-}_B +
\ket{+-}_A\ket{-+}_B\right). \nl{} \eqa Now Alice and Bob share a
mixture of two superpositions.  By Alice and Bob each measuring a
suitable observable (such as $\hat{O}=\uplus\ket{++}\uplus\ket{--}$
for example), they can perform a local measurement to discriminate
the two superposition states 
superpositions they actually
(without destroying the superposition). Thus they have access
to 1 ebit of constrained entanglement. In this case we started
with two copies of $\ket{\psi}$, therefore with a perfect
reference frame we would expect to be able to recover two ebits of
entanglement. However, even without an external reference frame it
is possible to access entanglement from two copies of the state.
This is because one of the states acts a reference for the other,
activating its entanglement. Alternatively, one could consider
that each of the entangled states acts a as partial reference
frame for the other, allowing half of its entanglement to be
accessed.  This is an example of a case where no entanglement
could be distilled from a single copy of the state (with no
reference frame), but two copies of the state allows entanglement
to be distilled.  Hence the state $\ket{\psi}$ is not 1-distillable, but
it is 2-distillable.  That is to say that this state demonstrates the
fact that the 1-distillable states are a subset of the 2-distillable states for the $S_2$-SSR.

\section{Beyond the $S_N$-SSR}

\subsection{Adding a stronger constraint}  So far we have
considered the problem of describing ensemble quantum information
processing using  the formalism for SSRs associated with some
group. 
The $S_N$-SSR says that all elements (molecules) are subject to
identical operations. This constraint has a demonstrable effect on
the properties of the system, which can however be removed through
use of additional resources such as reference frames.  We now wish
to consider the case where a stronger constraint than a SSR may
apply to a system.

First we point out a difference between NMR experiments and
spin-squeezing experiments, for which the $S_N$-SSR also applies.
In the latter, it is possible to perform symmetric operations
which entangle the elements (atoms), such as spin-squeezing
unitaries \cite{Ueda} or quantum non-demolition measurements of
$\hat{J}_z$ \cite{Kuz00}. By contrast, in NMR it is not possible
to induce correlations between different molecules. The reasons
for this difference are subtle, and relate to practical
constraints due to decoherence during the read-out. This
constraint also manifests itself in very low measurement
efficiencies, but here we ignore that issue.

Consider the $M=1$ case for simplicity. Then all that can be done
in practice in NMR experiments is
\begin{itemize}
\item \blk Rotations $\exp(-i\underline{\theta}\cdot\underline{\hat J}) = \exp(-i\underline{\theta}\cdot\sum_{k=1}^{N} \underline{\hat\sigma}^k/2)$.
\item Destructive measurement of $\hat J_z = \sum_{k=1}^{N} \hat\sigma_z^k /2$.
\end{itemize}
Here $\underline{\hat\sigma}^k$ denotes $I\otimes\ldots
I\otimes\underline{\hat\sigma}\otimes I\ldots\otimes I$ with
$\underline{\hat\sigma}$ in the $k$th position. When making a
measurement of this type (e.g. measuring $\hat J_z$) we actually
get out an overall signal which is proportional to the sum of the
spin ($\hat{\sigma}_z$) for each particle. Moreover, the final
state of the ensemble is unrelated to the measurement result, due
to thermal decoherence. Thus in general, the only operations
possible in NMR are to make destructive measurements of symmetric
observables that are additive over the ensemble: \beq\hat{O}_{\rm
total}=\sum_{k}\hat{O}^k, \label{NMROperations}\eeq where
$\hat{O}^k $ is the operator for the $k$th particle as above.
 We call such operations {\em non-collective}. This terminology is
 appropriate because the result of the measurement could be
 obtained by individually measuring each element of the ensemble
 and summing the results.

We can contrast such non-collective operations with a collective
operation like measuring (destructively or otherwise) $\hat{J}^2$
to find out the value of the total angular momentum $j$ for the
ensemble. This could not be done by measuring each particle and
summing the results. Previous work using the $S_N$-SSR assumed
that such collective measurements are possible. We will now
consider the case where operations need not only be
\emph{symmetric} but \emph{also non-collective}, as a stronger
constraint on the system.

We suspect that we cannot completely characterize these
constraints by any $G$-SSR.  Instead we must supplement the
$S_N$-SSR with the extra constraint that the operations also be
non-collective. This complicates matters, as we are now unable to
write down an equivalent state which is invariant under all the
allowable operations.  Despite this, we wish to determine if any
 entanglement survives under this stronger constraint.

Since we are unable to determine an operationally equivalent state
matrix for the constrained state we cannot calculate the
extractable entanglement directly. However, if a Bell inequality
violation can be demonstrated then this proves that entanglement
is present in some form. So the question becomes, using the
$S_N$-invariant state as a description for the system, is it
possible to demonstrate Bell nonlocality using non-collective
operations?

\subsection{Bell inequality for ensembles}
For specificity, we consider the problem of demonstrating Bell
nonlocality under symmetric, non-collective measurements on an
ensemble of $N=2J$ Bell singlets,
$\ket{\psi}=\ket{\psi^-}^{\otimes N}$. As discussed in Sec.
\ref{Loss} the interesting part of this state can be written for
simplicity as
\begin{equation}
\mathcal{P}\left[\left(\ket{\psi}\bra{\psi}\right)^{\otimes
N}\right]=\sum_{j=J-\lfloor
J\rfloor}^{J}{d_j}^2\wp_j|\phi_j\rangle\langle\phi_j|,\label{Ensemble}
\end{equation}
which is an incoherent mixture of different spin ($j$) states. The
added constraint means that we are unable to measure $\hat{J}^2$
directly, but can only measure components of spin (such as
$\hat{J}_z$). Thus we must be derive a Bell inequality that allows
for particles of different spin (i.e. different $j$ values).

Mermin \cite{Mer**} developed a Bell inequality for spin-$j$
particles by considering a generalization of the Bohm-EPR
experiment. The only assumption that needs to be satisfied for
this inequality to be applicable is that the desired state exhibit
perfect anticorrelation in the spins of the two particles. The
inequality can be written as, \bqa \left\langle\left|
m_A(\hat{a})- m_B(\hat{b})\right| \right\rangle &\geq&
\frac{1}{J}\Big(\langle m_A(\hat{a})m_B(\hat{c})\rangle
\Big.\nonumber\\ &&+\left.\langle
m_A(\hat{b})m_B(\hat{c})\rangle\right),\label{MInequality}\eqa
where $m_i(\hat{a})$ represents the spin component of the $i$th
particle in the $\hat{a}$ direction and $J$ is an upper bound on
the $m_i(\hat{a})$. For Mermin's case one can (and Mermin does)
choose $J=j$. However, we require that the parameter $J$ because
we cannot distinguish between different $j$-values. Inequality
(\ref{MInequality}) will be satisfied by any theory obeying local
causality.  For ease of analysis we define a quantity
\begin{eqnarray}M_J(\theta)&=& \left\langle\left|
m_A(\hat{a})-m_B(\hat{b})\right| \right\rangle\ \nonumber \\
&& -\frac{1}{J}\left(\langle
m_A(\hat{a})m_B(\hat{c})\rangle+\langle
m_A(\hat{b})m_B(\hat{c})\rangle\right).\label{MJ}\end{eqnarray}
The condition for local causality to be satisfied can thus be
expressed as $M_J(\theta)\geq 0$.

Consider a Stern-Gerlach experiment such that the spin can be
measured along one of three axes defined by coplanar vectors
$\hat{a}$, $\hat{b}$, and $\hat{c}$. Mermin defined these axes such
that the vectors $\hat{a}$ and $\hat{b}$ make the same angle
$\pi/2+\theta$ with $\hat{c}$, and the angle $\pi-2\theta$ with each
other. Using this set up for two perfectly anticorrelated spin-$j$
particles, quantum mechanics predicts that Eq. (\ref{MJ}) can be
expressed as \beq M_J^{\rm spin-j}(\theta) =
f_j(\theta)-\frac{1}{J}\frac{2j}{3}\left(j+1\right)\sin\theta,\label{Mj}\eeq
where the functions $f_j(\theta)$ are defined as \beq
f_j(\theta)=\frac{1}{2j+1}\sum_{m,m'}\left| m-m'\right| \left|
\bra{m}e^{-2i\theta \hat{S}_y}\ket{m'}\right|^2, \label{fj}\eeq
and $\hat{S}_y$ is a spin matrix.

Now an ensemble of Bell singlet states is perfectly anticorrelated
in spin and thus \erf{Ensemble} 
satisfies the necessary assumption for inequality
(\ref{MInequality}) to be applicable. Also, when Mermin evaluated
\erf{Mj} he assumed measurements of spin components, that is,
non-collective measurements. Thus it is possible to use the same
method as Mermin to evaluate the Bell inequality for an NMR
ensemble, as all the relevant constraints are accounted for. The
ensemble state simply behaves like an incoherent mixture of
different spin-$j$ states.

Thus, for an ensemble of Bell singlet states, quantum mechanics
predicts Eq. (\ref{MJ}) can be written as
\begin{equation} M_J^{\rm Ensemble}(\theta)= \sum_{j=J-\lfloor
J\rfloor}^Jd_j^2\wp_j M_J^{\rm spin-j}(\theta),\label{EnsembleMJ}
\end{equation} where $M_J^{\rm Ensemble}(\theta)<0$ demonstrates Bell-nonlocality.

\subsection{Demonstrating Bell nonlocality}

We are now in a position to show that Bell-nonlocality survives
under stronger constraints than those imposed by a SSR alone. To
do this we must evaluate $M_J^{\rm Ensemble}(\theta)$ and show
that it can become negative. To simplify this task it is
instructive to recall the form of Eq. (\ref{Mj}).  When Mermin
evaluated these terms, he found to a good approximation
(particularly for large $J$) that he was able to use a quadratic
form to simplify their evaluation. Using the same approximation
allows $M_J^{\rm Ensemble}(\theta)$ to be simplified to the
expression
\begin{equation}M_J^{\rm Ensemble'}(\theta)=\!\!\!\!\! \sum_{j=J-\lfloor
J\rfloor}^J\!\!\!\!\!d_j^2\wp_j\left[\frac{2}{3}
j\left(j+1\right)\sin\theta\left(2\sin\theta-\frac{1}{J}\right)\right]\!\!,\label{EnsembleApprox}
\end{equation} where the prime indicates an approximation.

Now, the probability terms $d_j^2\wp_j$ in Eq.
(\ref{EnsembleApprox}) are always positive, so the question
becomes, can the remaining factor be negative?  If this factor is
negative for all terms in the sum, then $M_J^{\rm
Ensemble'}(\theta)$ is negative and the \blk state exhibits Bell
nonlocality. Examining the terms in the sum more closely reveals
that there is always a linear (in $\sin\theta$) term subtracted
from a quadratic (in $\sin\theta$) term.  Hence, if $\theta$ (and
thus $\sin\theta$) is small enough, then the linear term will
always be dominant, resulting in a negative contribution to the
sum.  It is possible to choose $\theta$ to be small enough that
every term in the sum will be negative, thus $M_J^{\rm
Ensemble'}(\theta)<0$ and the ensemble state exhibits Bell
nonlocality.

To put it explicitly (by solving for $\theta$ in terms of $J$) the
ensemble state exhibits Bell-nonlocality despite the constraints
when the detectors can be arranged to make measurements defined by
$\theta$ where \begin{equation} 0 < \sin\theta <1/2J.\label{Range}
\end{equation}
This is actually a lower bound on the range of $\sin\theta$ for
which a violation is possible.  For small values of $J$ ($\leq3$),
Eq. (\ref{EnsembleMJ}) can be explicitly calculated (without
resorting to approximations). Even for these small values of $J$
the exact numerical results agree quite well \footnote{We expect
the approximation to work well for large $J$, however, even for
$J=1$ there is only 20\% difference between using the exact
evaluation of Eq. (\ref{EnsembleMJ}) and the approx. given by Eq.
(\ref{Range}). This difference drops to less than 8\% for $J=3$.
As $J\rightarrow\infty$ the difference between Eq.s
(\ref{EnsembleMJ}) and (\ref{Range}) vanishes.} with the range of
angles specified by Eq. (\ref{Range}) and the agreement improves
with larger $J$. This lends confidence that for large $J$ the
approximation leading to Eq. (\ref{Range}) is a valid one.

Somewhat surprisingly, Eq. (\ref{Range}) gives exactly the same
angular range for which Mermin demonstrated a pair of
(unconstrained) entangled spin-$J$ particles exhibit Bell
nonlocality. One may then ask which of the two systems, an
ensemble of Bell states or a pair of spin-$J$ particles, violates
the inequality more strongly.  A way to measure this is to
consider the depth of the violation, that is, how negative
$M_J(\theta)$ becomes.  For a pair of perfectly anticorrelated
spin-$J$ particles, the minimum value of $M_J(\theta)$ converges
to a constant value of $-1/12$ for large $J$. In contrast, for an
ensemble of $N=2J$ Bell states, the minimum of $M_J^{\rm
Ensemble'}(\theta)$ scales as $-1/J$. That is, the violation depth
tends to zero for large ensembles. Thus, a pair of spin-$J$
particles violates this Bell inequality more strongly than an
ensemble of $2J$ Bell states under our stronger constraint.

\section{Summary}
In this paper we have classified groups of states based on their
mixed state entanglement properties and related these states to
the well known concepts of activation and distillation. We have
also reviewed the analogy between mixed state entanglement and
that of pure state entanglement constrained by a SSR. In
particular we have focused on the symmetric group SSR. We have
demonstrated that the $S_N$-SSR limits the amount of entanglement
that can be accessed from an ensemble of entangled states. In
comparison with U(1)-SSRs such as the particle number SSR we show
how to apply the correct notion of multiple copies of an ensemble
state to asymptotically recover the entanglement lost due to the
SSR. We have also discussed the concepts of reference frames and
given examples to illustrate the similarities between concepts of
activation, distillation and use of reference frames (or multiple
copies of states) to recover entanglement. For the $S_2$-SSR we
showed that by using multiple copies of the ensemble, it is
possible to only lose 1 ebit of entanglement (asymptotically).

Finally we gave an example where it does not seem possible to
formulate the constraints on a system as a SSR. This situation
arises naturally in the context of a liquid NMR ensemble. The lack
of individual addressability requires that the $S_N$-SSR be
considered. However, other technical constraints arise due to the
large amount of thermal noise present in NMR ensembles. This noise
manifests itself in two ways: low measurement efficiency and the
fact that only non-collective measurements are possible.  We
addressed the latter manifestation and went on to show that
despite this stronger constraint it is still possible in principle
to demonstrate Bell nonlocality. It may prove interesting to
attempt also to include the effect of the low efficiency
constraint.

Further studies of physical constraints which cannot be formalized
as SSRs may prove a fruitful area of research, not only for
explaining experiments but also for understanding the properties
of entanglement in general.

\begin{acknowledgments}
This work was supported by the Australian Research Council and the
State of Queensland.  We thank P. Turner and A. C. Doherty for
useful discussions.

\end{acknowledgments}


\begin{thebibliography}{0}
\expandafter\ifx\csname natexlab\endcsname\relax\def\natexlab#1{#1}\fi
\expandafter\ifx\csname bibnamefont\endcsname\relax
  \def\bibnamefont#1{#1}\fi
\expandafter\ifx\csname bibfnamefont\endcsname\relax
  \def\bibfnamefont#1{#1}\fi
\expandafter\ifx\csname citenamefont\endcsname\relax
  \def\citenamefont#1{#1}\fi
\expandafter\ifx\csname url\endcsname\relax
  \def\url#1{\texttt{#1}}\fi
\expandafter\ifx\csname urlprefix\endcsname\relax\def\urlprefix{URL }\fi
\providecommand{\bibinfo}[2]{#2}
\providecommand{\eprint}[2][]{\url{#2}}

\end{thebibliography}


\begin{thebibliography}{99}\blk
\blk
\bibitem{NieChu00} \blk M. A. Nielsen and I. L. Chuang, \textit{Quantum
    Computation and Quantum Information} (Cambridge University Press,
  Cambridge, 2000).
\bibitem{Hor01} M. Horodecki, P. Horodecki and R. Horodecki, in
  \textit{Quantum Information: An Introduction to Basic Theoretical
  Concepts and Experiments}, Vol. 173 of \textit{Springer Tracts in
  Modern Physics} ed. G. Alber \textit{et al}. (Springer Verlag, Berlin,
  2001), pp. 151-195.
\bibitem{Barnum03} 
 H. Barnum, E. Knill, G. Ortiz, and L. Viola,
Phys. Rev. A {\bf 68}, 032308 (2003).
\bibitem{Barnum04}
H. Barnum, E. Knill, G. Ortiz, R. Somma, and L. Viola, Phys. Rev.
Lett. {\bf 92}, 107902 (2004).
\bibitem{KMP04} A. Kitaev, D. Mayers and J. Preskill, \pra
  \textbf{69}, 052326 (2004).
\bibitem{BarWis03} S. D. Bartlett and H. M. Wiseman, \prl \textbf{91},
  097903 (2003).

\bibitem{VacAnsWisJac06}  
J. A. Vaccaro, F. Anselmi, H. M. Wiseman, and K. Jacobs,
quant-ph/0501121 (unpublished).
\bibitem{BarDohSpeWis06}
S. D. Bartlett, A. C. Doherty, R.W. Spekkens, and H. M. Wiseman,
Phys. Rev. A {\bf 73}, 022311 (2006).

\bibitem{Ver03} F. Verstraete and J. I. Cirac, \prl \textbf{91},
  010404 (2003).
\bibitem{WisVac03} H. M. Wiseman and J. A. Vaccaro, \prl, \textbf{91},
  097902 (2003).

 \bibitem{SchVerCir04}
N. Schuch,  F. Verstraete, and J. I. Cirac,
 Phys. Rev. A {\bf 70}, 042310 (2004).

\bibitem{Wis04}
H. M. Wiseman,
 J. Opt. B. {\bf 6}, S849-S859 (2004).
\bibitem{SanBarRudKni03}
B. C. Sanders, S. D. Bartlett, T. Rudolph, and P. L. Knight,
Phys. Rev. A {\bf 68}, 042329 (2003).
 \bibitem{RudSan01}
T. Rudolph and B. C. Sanders,  Phys. Rev. Lett. {\bf 87}, 077903
(2001).

\bibitem{EnkFuc02a}
S. J. van Enk  and C. A. Fuchs,
Phys. Rev. Lett. {\bf 88}, 027902 (2002).

\bibitem{vKK04} J. von Korff and J. Kempe, \prl
\textbf{93}, 260502 (2004). %
\bibitem{Bel64}
J. S. Bell, 
Physics (New York) {\bf 1}, 195 (1964).
\bibitem{Sch35}
E. Schr\"{o}dinger, Naturwissenschaften {\bf 23}, 807 (1935);
English translation by J. D. Trimmer in Proc. Am. Philosophical
Soc. {\bf 124}, 323 (1980).
\bibitem{Sch36} 
E. Schr\"odinger,
 Proc. Camb. Phil. Soc. {\bf 32}, 446  (1936). 
\bibitem{Wer89} R. F. Werner, \pra \textbf{40}, 4277
  (1989).
\bibitem{Ben96} C. H. Bennett \textit{et al.},
  \prl \textbf{76}, 722 (1996); C. H. Bennett, 
 D. P. DiVincenzo, J. A. Smolin and W. K. Wootters,
  \pra \textbf{54}, 3824 (1996).
\bibitem{Hor98} M. Horodecki, P. Horodecki and R. Horodecki, \prl
  \textbf{80}, 5239 (1998).
\bibitem{Gur02}
L. Gurvits, quant-ph/0201022 (unpublished).
\bibitem{DohParSpe02}
A. C. Doherty, P. A. Parrilo, and F. M. Spedalieri, Phys. Rev.
Lett. {\bf 88}, 187904 (2002).
\bibitem{Div00} D. P. DiVincenzo, 
P. W. Shor, J. A. Smolin, B. M. Terhal and A. V. Thapliyal,
  \pra \textbf{61}, 062312 (2000).

\bibitem{Dur00} W. D\"ur, J. I. Cirac, M. Lewenstein and D. Bru{\ss},
  \pra \textbf{61}, 062313 (2000).

\bibitem {Hor97} M. Horodecki, P. Horodecki and R. Horodecki, \prl
  \textbf{78}, 574 (1997).

\bibitem{Wat04} J. Watrous, \prl \textbf{93}, 010502
  (2004).



\bibitem{Rai01} E. M. Rains, IEEE Trans. Inf. Theory \textbf{47}, 2921
  (2001).

\bibitem{Egg01} T. Eggeling, KGH. Vollbrecht, R. F. Werner, and M. M. Wolf,
\prl \textbf{87}, 257902 (2001).
\bibitem{Hor99} P. Horodecki, M. Horodecki and R. Horodecki, \prl
  \textbf{82}, 1056 (1999).
\bibitem{HillWoot97}
S. Hill and W. K. Wootters, Phys. Rev. Lett. {\bf 78}, 5022
(1997).
\bibitem{BennetDiVSmolWoot96}
C. H. Bennett, D. P. DiVincenzo, J. A. Smolin, and W. K. Wootters,
Phys. Rev. A {\bf 54}, 3824 (1996).
\bibitem{Wic52}
G. C. Wick, A. S. Wightman, and E. P. Wigner, Phys. Rev. {\bf 88},
101 (1952).

\bibitem{AhaSus67}
Y. Aharonov and L. Susskind, Phys. Rev. {\bf 155}, 1428 (1967).
\bibitem {BarRudSpe05} S. D. Bartlett, T. Rudolph and R. W. Spekkens,
 Int. J. Quantum Inf. {\bf 4}, No. 1, 17 (2006).
\bibitem{OppHor02}
J. Oppenheim, M. Horodecki, P. Horodecki, and R. Horodecki, Phys.
Rev. Lett. {\bf 89}, 180402 (2002).

\bibitem{Kni00} E. Knill, R. Laflamme and L. Viola, \prl \textbf{84},
  2525 (2000).
\bibitem{NMRQIP} D. G. Cory \emph{et al.}, Proc. Natl. Acad. Sci. U.S.A. {\bf 94}, 1634
(1997); N. Gershenfeld and I. L. Chuang, Science {\bf 275}, 350
(1997).

\bibitem{Ueda}
M. Kitagawa and M. Ueda, Phys. Rev. A {\bf 47}, 5138 (1993).

\bibitem{Anw04}
M. S. Anwar \textit{et al.}, Phys. Rev. Lett. {\bf 93}, 040501
(2004)

\bibitem {Ful91} W. Fulton and J. Harris, \textit{Representation
    theory: a first course}, (Springer-Verlag, Berlin, 1991).

\bibitem{Eis00}
J. Eisert, T. Felbinger, P. Papadopoulos, M. B. Plenio, and M.
Wilkens, Phys. Rev. Lett. {\bf 84}, 1611 (2000).

\bibitem{Herbut03}
F. Herbut, J. Phys. A: Math. Gen. {\bf 36} 8479 (2003).

\bibitem{Peres96}
A. Peres, Phys. Rev. Lett. {\bf 77}, 1413 (1996).

\bibitem{Kuz00}
A. Kuzmich, L. Mandel, and N. P. Bigelow,  Phys. Rev. Lett. {\bf
85}, 1594 (2000).

\bibitem{Mer**}
N. D. Mermin, Phys. Rev. D {\bf 22}, 356 (1980). 

\end{thebibliography}
\end{document}